\documentclass[
preprint,
aps,
prd,
showpacs,
nofootinbib,
superscriptaddress,
tightenlines,
floatfix]{revtex4-1}

\usepackage{amsmath}
\usepackage[dvips]{graphicx}
\usepackage{epstopdf}
\usepackage{epsfig}
\usepackage{amssymb}
\usepackage{bm}
\usepackage{slashed}
\usepackage{color}

\newcommand{\beq}{\begin{equation}}
\newcommand{\eeq}{\end{equation}}
\newcommand{\ov}{\overline}
\newcommand{\pa}{\partial}

\begin{document}

\vbox{
\baselineskip 14pt
\hfill \hbox{\rm{\footnotesize KUNS-2563}}\\
\hfill \hbox{\rm{\footnotesize EPHOU-15-010}}\\
\hfill \hbox{\rm{\footnotesize YITP-15-43}}
} \vskip 1.7cm

\title{
\bf Study of lepton flavor violation
\\ in flavor symmetric models for lepton sector}

\author{Tatsuo Kobayashi}
\affiliation{ Department of Physics, Hokkaido University, Sapporo 060-0810, Japan}

\author{Yuji Omura}
\affiliation{Kobayashi-Maskawa Institute for the Origin of Particles and the Universe (KMI), Nagoya University, Nagoya 464-8602, Japan}

\author{Fumihiro Takayama}
\affiliation{Yukawa Institute for Theoretical Physics, Kyoto University, Kyoto 606-8502, Japan}

\author{Daiki Yasuhara}
\affiliation{Department of Physics, Kyoto University, Kyoto 606-8502, Japan}

\begin{abstract} 
Flavor symmetric model is one of the attractive Beyond Standard Models (BSMs) 
to reveal the flavor structure of the Standard Model (SM). 
A lot of efforts have been put into the model building 
and we find many kinds of flavor symmetries and setups are able 
to explain the observed fermion mass matrices. 
In this paper, we look for common predictions of physical observables 
among the ones in flavor symmetric models, 
and try to understand how to test flavor symmetry in experiments. 
Especially, we focus on the BSMs for leptons 
with extra Higgs $SU(2)_L$ doublets charged under flavor symmetry. 
In many flavor models for leptons, remnant symmetry is partially respected 
after the flavor symmetry breaking, 
and it controls well the Flavor Changing Neutral Currents (FCNCs) 
and suggests some crucial predictions against the flavor changing process, 
although the remnant symmetry is not respected in the full lagrangian. 
In fact, we see that 
$\tau^- \to e^+ \mu^- \mu^-$ $( \mu^+ e^- e^-)$ 
and $e^+ e^- \to \tau^+\tau^-$ $(\mu^-\mu^+)$ 
processes are the most important in the flavor models 
that the extra Higgs doublets belong to triplet representation 
of flavor symmetry. For instance,
the stringent constraint from the $\mu \to e \gamma$ process could be evaded 
according to the partial remnant symmetry.
We also investigate the breaking effect of the remnant symmetry 
mediated by the Higgs scalars, 
and investigate the constraints from the flavor physics: 
the flavor violating $\tau$ and $\mu$ decays, the electric dipole moments, 
and the muon anomalous magnetic moment. 
We also discuss the correlation between FCNCs and nonzero $\theta_{13}$, 
and point out the physical observables in the charged lepton sector 
to test the BSMs for the neutrino mixing.
\end{abstract}

\maketitle

\section{Introduction}
\label{sec1}

We know that there are three generations of fermions in nature. 
Each generation carries the same quantum number, 
and only their masses are different from each other. 
In the Standard Model (SM), the transition among the generations occurs 
only through the weak boson exchanging, but the flavor-changing processes 
via the Flavor Changing Neutral Currents (FCNCs) are strongly suppressed 
because of the Glashow-Iliopoulos-Maiani (GIM) mechanism \cite{Glashow:1970gm}. 
This SM picture successfully describes the experimental results, 
but we may wonder why such a flavor structure exists in our nature. 
We expect that Beyond Standard Model (BSM) exists 
out of our current experimental reach, 
and reveals the origin of the three generations. 
One of the promising BSMs is a flavor symmetric model. 

In the SM, flavor symmetry is explicitly broken by Yukawa couplings 
to generate fermion mass matrices. 
Without the couplings, we could find $SU(3)$ symmetry in each sector of 
left-handed (right-handed) up-type, down-type quarks and leptons. 
In flavor symmetric models, 
the $SU(3)$ symmetry or the subgroup of $SU(3)$ is respected in the Lagrangian, 
introducing extra scalar bosons charged under the flavor symmetry. 
For instance, additional $SU(2)_L$-doublet Higgs fields are introduced, 
and the scalars and fermions are charged under the flavor symmetry 
to write down Yukawa couplings. 
The flavor-charged Higgs fields develop nonzero vacuum expectation values (VEVs), 
and break not only the electro-weak (EW) symmetry but also the flavor symmetry. 
Then the flavor structure of the SM is 
effectively generated at the low-energy scale. 
This BSM would be very attractive and reasonable, 
and so many types of flavor symmetric models have been proposed so far \cite{Altarelli,Ishimori,S4}. 
Especially, we could find so many models motivated by the large mixing 
in neutrino sector, because the experimental result may imply so-called 
Tri-Bi maximal mixing \cite{TB}, which can be easily accommodated by the BSMs 
with non-Abelian discrete flavor symmetry such as 
$A_4$ \cite{A4,A4-3,flavor2}, 
$S_4$ \cite{S4-0}, 
and $\Delta (27)$ \cite{delta27-1} {\it etc}..
Recent result on the nonzero $\theta_{13}$ 
\cite{theta13-1,theta13-2,theta13-3,theta13-4,theta13-5} 
may require some small modifications 
in those models, but we could expect that 
so many kinds of flavor symmetric models 
can be still consistent with the experimental results 
\cite{A4-2,A4-4,Hamada:2014xha,S4,S4-3,delta27,flavor-higher,theta13-model}. 
Then, the next question in this approach to the flavor structure would be 
how to test the flavor symmetry in experiments. 

One hint to clarify which kinds of symmetry exist behind the flavor structure 
would be obtained, if we consider the origin of the remnant symmetry 
in the fermion mass matrices in the SM. 
As we discuss in Sec. \ref{sec2}, we see that there is symmetry in 
mass matrices of leptons in the SM, which is explicitly broken 
by the weak interaction involving $W$ boson. 
If flavor symmetry exists behind the SM, 
the remnant symmetry might be the fragment of the flavor symmetry 
broken at high energy. In fact, one can find a lot of works on flavor models, 
based on the assumption that the symmetry in the fermion mass matrices is 
the subgroup of flavor symmetry \cite{King,Altarelli,Ishimori,S4,S4-3}. 

In this paper, we investigate especially FCNCs 
involving charged leptons in flavor symmetric models, where 
the symmetry in the charged lepton mass matrix is the subgroup 
of the flavor symmetry and extra $SU(2)_L$-doublet Higgs fields 
charged under the additional symmetry are introduced. 
Once we assume that the symmetry is originated from 
the flavor symmetry spontaneously broken at high energy, 
we find that the FCNCs involving the extra Higgs fields are predicted 
by the remnant symmetry. 
The remnant symmetry would not be respected in the full Lagrangian, 
but it could well control the FCNCs as long as 
the breaking terms are enough small in the Higgs potential. 
In Sec. \ref{sec2}, we discuss the remnant symmetry and our setup in this paper. 
Then we investigate the FCNCs involving neutral scalars and charged leptons, 
and discuss Higgs potential in Sec. \ref{section3}. 
We see that the partially remnant symmetry in the lepton mass matrix 
well controls FCNCs, and study our signals and 
current experimental constraints in flavor physics in Sec. \ref{section4}. 
On the other hand, it is also one of crucial issues to understand 
how to derive nonzero $\theta_{13}$ in flavor symmetric models. 
As we mentioned above, nonzero $\theta_{13}$ may require some modifications 
in conventional setups, because simple scenarios tend to predict 
vanishing $\theta_{13}$. We study the correlation between $\theta_{13}$, 
especially given by the mixing angles in charged lepton sector, and FCNCs, 
in Sec. \ref{section5}. Section \ref{section6} is devoted to summary. 
In the Appendix A, we introduce the $A_4$ model as a concrete example.

\section{Generic Argument about FCNCs in flavor models}
\label{sec2}

In the SM, the fermions obtain masses 
according to the nonzero VEV of a Higgs field and Yukawa couplings. 
The Yukawa couplings should be defined to realize the large mass hierarchies 
and mixing in quarks and lepton sectors, 
so that the flavor symmetry that rotates the flavors 
is explicitly broken by the couplings in the SM. 
Furthermore, the charged currents involving $W$ boson change the flavors, 
so that it would be difficult to find out even very simple flavor symmetry 
such as $Z_2$ and $Z_3$ in the full Lagrangian of the SM.

Now, let us focus on mass matrices of leptons and look for symmetry 
that is respected in only each mass matrix. 
For instance, if we see only the mass matrices 
for the charged lepton $(M_l)$ and the neutrinos $(M_N)$, 
we could find flavor symmetry as, 
\beq \label{remnant}
M_l M^\dagger_l =T_L M_l M^\dagger_l T^\dagger_L, ~  M^\dagger_l M_l =T_R M^\dagger_l M_l T^\dagger_R, ~ M_N=S^T M_N S,
\eeq
assuming neutrinos are Majorana particles (See Ref. \cite{S4}, for instance). 
When $M_l$ and $M_N$ are diagonal, $T_L$, $T_R$, and $S$ could be described as 
\beq
T_L=\begin{pmatrix} 1 & 0 & 0 \\ 0 & \eta_L & 0 \\ 0 & 0 & \eta_L^* \end{pmatrix},~ T_R=\begin{pmatrix} 1 & 0 & 0 \\ 0 & \eta_R & 0 \\ 0 & 0 & \eta_R^* \end{pmatrix},~ S=\begin{pmatrix} (-1)^p & 0 & 0 \\ 0 & (-1)^q & 0 \\ 0 & 0 & (-1)^{p+q} \end{pmatrix}, 
\eeq
where $\eta_L$ and $\eta_R$ are the complex numbers which satisfy 
$\eta_L \eta_L^*=\eta_R \eta_R^*=1$, and $p, \, q$ are integer. 
$T_L$, $T_R$, and $S$ are not conserved in the full Lagrangian. 
In fact, they are broken by the gauge interaction with $W$ boson explicitly. 
However, they may give a hint for the mystery of the flavor structure in the SM. 
As discussed in Refs. \cite{King,Altarelli,Ishimori,S4,S4-3}, 
we can find the remnant symmetry, $T_L$, $T_R$, and $S$ in the flavor models 
which explain the realistic mass matrices naturally, 
and the remnant ones could be interpreted as the subgroup 
of the original flavor symmetry. 
Below, let us discuss such a kind of flavor models 
and $SU(2)_L$-doublet extra Higgs fields charged under the flavor symmetry, 
and consider the scenario that the simple remnant symmetry appears 
after the symmetry breaking. 

\subsection{Remnant symmetry in flavor symmetric models}
\label{sec2-1}

Let us consider flavor symmetric models with extra Higgs doublets. 
The extra symmetry may be non-Abelian discrete symmetry, 
and the extra scalars may belong to non-trivial singlet, doublet or triplet. 
Let us focus on Yukawa couplings of charged lepton sector 
in flavor symmetric models. 
In general, the couplings for the fermion masses could be described as 
\beq
{\cal L}=-\overline{L}_i M_l(\phi)^{ij} E_{R \, j}+h.c..
\eeq
$\phi$ is a scalar charged under the EW symmetry and the flavor symmetry, 
and we simply assume that $M_l$ only depends on $\phi$. 
Then $M_l$ satisfies the following relation 
according to the flavor symmetry $({\cal G})$ with generators $(g)$; 
\beq
M_l( \phi)=g_L M_l( g^\dagger_{\phi} \phi) g_R.
\eeq
$g_{ L, \, R, \, \phi}$ are defined corresponding to the representations of 
$L_i$, $E_{R \, i}$, and $\phi$ under ${\cal G}$. 
When $\phi$ develops the nonzero VEV, the EW and flavor symmetry are broken 
and mass matrix for charged leptons is generated. Let us simply assume that 
the remnant symmetry of the flavor symmetry (${\cal T}$), 
whose generator is $T \subset g$, 
is still hold in the mass matrix as follows: 
\begin{eqnarray}
\label{remnant2-1}
M_l (\langle \phi \rangle ) &=&T_L M_l   (T^\dagger_{\phi}\langle \phi \rangle )T^\dagger_R=T_L M_l   (\langle \phi \rangle )T^\dagger_R, \\
T^\dagger_{\phi}\langle \phi \rangle &=&T_{\phi}\langle \phi \rangle=\langle \phi \rangle.
\label{remnant2}
\end{eqnarray}
Let us consider the case that $L_i$ is triplet-representation 
in the diagonal base of $T_L$. Then $T_L$ for $L_i$ would be, 
\beq\label{T}
T_L=\begin{pmatrix} 1 & 0 & 0 \\ 0 & \eta_L & 0 \\ 0 & 0 & \eta_L^* \end{pmatrix}.
\eeq
If $\eta_L$ is not $\pm 1$, we find that $L_i$ in the diagonal base of $T_L$ is 
identical to the field ($l_i$) in the mass base 
according to the relation of Eq. (\ref{remnant}). 
In this paper, we only focus on the case with $\eta_L \neq \pm 1$. 
Moreover, we especially discuss flavor models 
with flavor triplet-representation Higgs doublet $(\phi \equiv H_i)$, 
so that the VEV alignment is given by Eqs. (\ref{remnant2}) 
and (\ref{T}) with $T_L=T_\phi$: 
\beq
(\langle \phi_1 \rangle, \, \langle \phi_2 \rangle, \, \langle \phi_3 \rangle) \propto (1, \, 0, \, 0).
\eeq
The orthogonal directions would be in the mass base of scalars around the VEV, 
and they may also respect the remnant symmetry, ${\cal T}$. 
Furthermore, the mass base of $E_{R \, i}$ is also fixed 
by $T_R$ as we discuss below, 
so that we can expect that it is possible that 
the FCNCs involving Higgs fields are qualitatively discussed 
in this kind of scenario, not mentioning original symmetry ${\cal G}$. 

\subsection{Setup}

Below, we focus on flavor models 
with triplet-representation Higgs doublet $H_i$. 
The Yukawa coupling for charged lepton is given by 
\beq
{\cal L}=-\overline{L}_i \Hat{Y}^k_{ij} H_j E_R^k+h.c..
\eeq
The texture of the matrices, $\Hat{Y}^k_{ij}$, is fixed by ${\cal G}$, 
and we can find this type of setups in Refs. \cite{A4,S4,T7,delta27,flavor}. 
\footnote{$SU(2)_L$ singlets charged under flavor symmetry are introduced 
allowing higher-dimensional operators in Refs. \cite{A4-3,flavor-higher}. 
Such kind of models are not be considered in this paper, 
but could be also related to our studies.}
Our assumptions of our setup are as follows: 
\begin{itemize}
\item 
$L_i$ and $H_i$ are triplet representations of ${\cal G}$,
\item
$\langle H_i \rangle$ breaks ${\cal G}$ to ${\cal T}$,
\item
$E_{R \, i}$ is non-trivial singlet of ${\cal T}$.  
\end{itemize}
${\cal T}$ would not be conserved in the full Lagrangian 
but partially respected, i.e. in the charged lepton Yukawa couplings. 
As discussed in subsection \ref{sec2-1}, 
charged lepton mass matrices only hold the symmetry as in Eq. (\ref{remnant2-1}). 
It is well-known that this situation successfully realizes 
realistic mass matrices according to the Tri-Bi maximal mixing structure 
in flavor models with non-Abelian discrete symmetry: 
for instance $A_4$ \cite{A4,A4-3}, $S_4$ \cite{S4,S4-0, A4-2,S4-2,S4-3}, 
$A_5$ \cite{A5}, $T_7$ \cite{T7}, $\Delta(27)$ \cite{delta27}, 
and $\Delta(6n^2)$ \cite{delta6n2}. 
Note that $E_{R \, i}$ may belong to the 
triplet-representation or non-trivial singlet of ${\cal G}$ 
before the symmetry breaking, but we do not specify it. 

\subsection{Mass base of charged leptons}

$L_i$ and $H_i$ are triplet-representation of ${\cal G}$, 
so that they are also the triplet of ${\cal T}$. 
Let us denote $L_i$ by the fields $(l_i)$ in the diagonal base of $T_L$. 
Then $l_i$ are in the mass base as we discussed above. 
Let us denote $E_{R \, i}$ as $e_{R \, i}$ in this base. 
Then, $e_{R \, i}$ are also the ones in the mass base, which transform as 
$(e_{R \, 1}, \, e_{R \, 2}, \, e_{R \, 3}) \to (e_{R \, 1}, \, \eta_L e_{R \, 2}, \, \eta_L^* e_{R \, 3})$,
because of Eq. (\ref{remnant2-1}). 
Eventually, the texture of $\Hat Y^k_{ij}$ is almost fixed 
because of $T_L$ in Eq. (\ref{T}): 
\beq
(\Hat Y^1_{ij}) = \frac{\sqrt 2}{v \cos \beta}\begin{pmatrix} m_1 & 0 & 0 \\ 0 & b_1 & 0 \\ 0 & 0 & c_1 \end{pmatrix},
\, (\Hat Y^2_{ij}) =\frac{\sqrt 2}{v \cos \beta} \begin{pmatrix}0 & 0 & c_2 \\ m_2 & 0 & 0 \\ 0 & b_2 & 0 \end{pmatrix},
\, (\Hat Y^3_{ij}) = \frac{\sqrt 2}{v \cos \beta} \begin{pmatrix} 0 & b_3 & 0 \\ 0 & 0 & c_3 \\ m_3 & 0 & 0 \end{pmatrix},
\eeq
where, $\langle H_1 \rangle=v\cos\beta/\sqrt{2}$. 
See Eq. (\ref{eq:betaisdefinedhere}).
Nonzero $b_2$ and $c_3$ imply ${\cal T}=Z_3$.\footnote{In our analysis, we assume the relation of Eq. (\ref{A4-relation}),
so ${\cal T}$ is set to $Z_3$. }
If $\Hat Y^k_{ij}$ is given by $\Hat Y^k_{ij}=y^k S^k_{ij}$, 
where $S^k_{ij}$ is defined by the multiplication rule of ${\cal G}$ 
and $y^k$ are dimensionless couplings, 
the elements of $\Hat Y^k_{ij}$ could be estimated, substituting 
\beq
\label{A4-relation}
|b_i|=|c_i|=m_i,
\eeq
where $m_i$ are the charged lepton masses. 
In this case, the mass matrix for charged lepton ($(M_l)^k_i$) is given by 
\beq
(M_l)^k_i=\frac{v \cos \beta}{\sqrt 2} \Hat Y^k_{i 1}.
\eeq
Below, we discuss flavor physics assuming the relation 
in Eq. (\ref{A4-relation}). 

\subsection{Mass base of scalars}

After the EW and flavor symmetry breaking, we find several scalars: 
CP-even, CP-odd, and charged scalars. 
In addition to flavor-triplet $H_i$, 
we introduce one flavor-singlet Higgs field, $H_q$, 
in order to realize the realistic mass matrices for quarks. 
We may need another flavor-charged scalars, $\Phi$, 
to generate Majorana mass matrices for neutrino mixing and masses. 
$\Phi$ may break the subgroup ${\cal T}$, 
and the mixing between $H_i$ and $\Phi$ may be allowed in the lagrangian. 
The mixing term may break the vacuum alignment 
as discussed in Eq. (\ref{remnant2}), 
because the VEV of $\Phi$ corresponds to the ${\cal T}$ breaking term. 
Below, we give some discussion about the mixing, 
and let us study $H_i$ and $H_q$, first. 

Let us decompose the scalars as follows: 
\beq
\label{eq:betaisdefinedhere}
H_q= \begin{pmatrix} H^{+}_q \\  \frac{1}{\sqrt 2} (v \sin \beta+H^0_q + i \chi_q) \end{pmatrix}, ~ H_1= \begin{pmatrix} H^{+}_1 \\  \frac{1}{\sqrt 2} (v\cos \beta +H^0_1 + i \chi_1) \end{pmatrix},
\eeq
and
\beq
H_2=\begin{pmatrix} H^{+}_{e} \\ \phi_{e} \end{pmatrix},~H_3=\begin{pmatrix} H^{+}_{\mu} \\ \phi_{\mu} \end{pmatrix}. 
\eeq
$H_q$ and $H_1$ generally mix each other because they develop nonzero VEVs: 
%-----------------
\beq
\begin{pmatrix} H_q^+ \\ H_1^+ \end{pmatrix} = \begin{pmatrix} \cos \beta \\ \sin \beta \end{pmatrix}G^+ + 
\begin{pmatrix} -\sin \beta \\ \cos \beta \end{pmatrix}H_S^+,
\label{mixinghp}%
\eeq
%-----------------
%-----------------
\beq
\begin{pmatrix} \chi_q \\ \chi_1 \end{pmatrix} = \begin{pmatrix} \cos \beta \\ \sin \beta \end{pmatrix}G^0 + 
\begin{pmatrix} -\sin \beta \\ \cos \beta \end{pmatrix}A_S,
\label{mixinghp}%
\eeq
%-----------------
%-----------------
\beq
\begin{pmatrix} H^0_q \\ H^0_1 \end{pmatrix} = \begin{pmatrix} \cos \alpha \\ \sin \alpha \end{pmatrix}H^0_{S \, 1} + 
\begin{pmatrix} -\sin \alpha \\ \cos \alpha \end{pmatrix}H^0_{S \, 2}.
\label{mixinghp}%
\eeq
%-----------------
$G^0$ and $G^+$ are the Goldstone boson eaten by $Z$ and $W^+$ bosons. 
If ${\cal T}$ is conserved in the mass matrices of scalars, 
$H_S^+$, $A_S^+$, $H^0_{S \, 1}$, and $H^0_{S \,2}$ are in the mass bases, 
and they do not mix with $H_2$ and $H_3$ for the ${\cal T}$ charge conservation. 
We will give some discussions about the mixing in Sec. \ref{section3}. 

$\alpha$ is the mixing angle between two CP-even scalars, 
and fixed by Higgs potential. 
If we build Higgs potential to lead SM-like Higgs mass and signal strength, 
$\alpha$ should be identical to $\beta$, 
and $H^0_{S \, 1}$ is interrupted as the SM Higgs.

On the other hand, $H^+_e$, $H^+_\mu$, $\phi_e$ and $\phi_\mu$ 
are the complex scalars to carry the ${\cal T}$ charges: 
$H_2 \to \eta H_2$ and  $H_3 \to \eta^* H_3$. 
In general, $H_2$ and $H_3$ would mix each other according to the nonzero VEV 
$\langle \Phi \rangle$, 
because $\langle \Phi \rangle$ breaks ${\cal T}$ spontaneously. 
We discuss the effect against the observables in  flavor physics later. 

\subsection{Yukawa couplings}
\label{sec2-E}

Now we define ${\cal T}$-conserving Yukawa couplings involving scalars. 
Based on the above argument, we find the following Yukawa couplings 
which induce flavor violations: 
\beq 
{\cal L_T}=-Y^{ij}_{e} \phi_e \overline{l}_i e_{R \, j} -Y^{ij}_{\mu} \phi_\mu \overline{l}_i e_{R \, j} -(V^\dagger)_{ik} Y^{kj}_{e} H^+_e \overline{\nu_L}_i e_{R \, j} -(V^\dagger)_{ik}Y^{kj}_{\mu} H^+_\mu \overline{\nu_L}_i e_{R \, j}+h.c.,
\eeq
where $V$ is the PMNS matrix. $Y^{ij}_{e}$ and $Y^{ij}_{\mu}$ are defined as
\beq
\label{eq:Yukawa}
(Y^{ij}_{e})=\Hat Y^j_{i2}=\frac{\sqrt 2}{v \cos \beta} \begin{pmatrix} 0 & 0 & b_3 \\  b_1 & 0 & 0 \\ 0 & b_2 & 0 \end{pmatrix},~
(Y^{ij}_{\mu})=\Hat Y^j_{i3}=\frac{\sqrt 2}{v \cos \beta} \begin{pmatrix} 0 & c_2 & 0 \\  0 & 0 & c_3 \\ c_1 & 0 & 0 \end{pmatrix}.
\eeq
As we mentioned above, the complex scalars may not be in the mass bases, 
because of ${\cal T}$-breaking effects in the Higgs potential. 
In Sec. \ref{section4}, we investigate the FCNC contributions to flavor physics 
in the ${\cal T}$-conserving limit, and then discuss the corrections 
from the ${\cal T}$-breaking terms in the Higgs potential 
to the observables in flavor physics. 
In fact, the breaking effect is strongly constrained 
by the $\mu \to e \gamma$ process. 

On the other hand, 
the neutral and charged scalars from $H_q$ and $H_1$ 
consist of Yukawa couplings that are the same as 
the model called type-X 2HDM in \cite{Aoki:2009ha}, 
or lepton-specific 2HDM in \cite{Branco:2011iw}: 
\begin{eqnarray}
{\cal L}_{\rm 2HDM}&=&-\frac{m_i \sin \alpha}{v \cos \beta} H^0_{S \, 1} \overline{l}_i e_{R \, j}  -\frac{m_i \cos \alpha}{v \cos \beta} H^0_{S \, 2} \overline{l}_i e_{R \, j} +h.c.   \nonumber  \\
&-&i\frac{m_i}{v \tan \beta} A_S \overline{l}_i e_{R \, i} -V_{ij}\frac{m_i}{v \tan \beta} H^+_S \overline{\nu_L}_i e_{R \, j} +h.c..
\label{2HDM}
\end{eqnarray}
The phenomenology of lepton-specific 2HDMs has been studied well 
in Refs. \cite{Aoki:2009ha, Branco:2011iw}. 

\section{Study of the Higgs potential}
\label{section3}

Before studying the phenomenological aspects, 
let us discuss Higgs potential in flavor symmetric models. 
In our setup, flavor-triplet $H_i$ develops nonzero VEV in the direction of 
$(\langle H_1 \rangle, \, \langle H_2 \rangle, \, \langle H_3 \rangle) \propto (1, \, 0, \, 0)$, 
and ${\cal T}$ is not broken. 
$\Phi_i$ is $SU(2)_L$-singlet and breaks ${\cal G}$ 
to the subgroup ${\cal S}$ of ${\cal G}$. 
In general, ${\cal S}$ and ${\cal T}$ are not commutative, 
so that $\langle \Phi_i \rangle$ breaks ${\cal T}$ 
and how to realize the vacuum alignment may be one of the issues in our models. 
For instance, the mechanism to achieve rigid vacuum alignment 
has been proposed so far \cite{Kobayashi:2008ih}.

In order to realize the vacuum alignment that respects ${\cal T}$, 
especially mixing term between $H_i$ and $\Phi_i$ should be controlled. 
In general, the Higgs potential is written as 
\beq
\label{potential}
V= V_H(H_q, \, H_i) + V_{\Phi} (\Phi_i, \, H_q) + \Delta V (\Phi_i, \, H_i, \, H_q), 
\eeq 
where $\Delta V$ only has the mixing terms between $\Phi_i$ and $H_i$ 
such as $|H^i|^2 |\Phi_i|^2$ and $ H_i^\dagger H_q \Phi_i$. 
If $\Delta V$ is absent, 
the vacuum alignment of $\langle H_i \rangle$ and $\langle \Phi_i \rangle$ 
are independently fixed by $V_H$ and $V_{\Phi}$. 
In this case, the mass matrices for the scalars originated from $H_i$ and $H_q$ 
would respect ${\cal T}$-symmetry, 
while the mass matrices from $H_q$ and $\Phi_i$ 
would respect ${\cal S}$-symmetry. 
This means that scalar mass eigenstates are decided 
according to only the remnant symmetries in each sector, 
and the flavor violating Yukawa couplings in the mass base of scalars 
are given by Eq. (\ref{eq:Yukawa}). 

We consider an example to illustrate our argument. 
In the absence of $\Delta V$, we can write down the mass matrix 
for the CP-even scalar mass eigenstates 
after the spontaneous symmetry breaking. 
On the basis of 
$(H_q^0, H_1^0, H_e^0, H_\mu^0, \Phi_1^0, \Phi_2^0, \Phi_3^0)^T$, 
where $\phi_e=\frac{1}{\sqrt{2}}(H^0_e+iA_e)$, 
$\phi_\mu=\frac{1}{\sqrt{2}}(H^0_\mu+iA_\mu)$ and 
$\Phi_j=\frac{1}{\sqrt{2}}(v_{\Phi}+\Phi_j^0+iA_j^\Phi)$ 
are defined, the mass matrix of the CP-even mass scalars 
is given in the following form: 
\beq
\left(
\begin{array}{c|cc}
M^2_q      & m^{2T}_{H} & m^{2T}_{\Phi} \\\hline
m^2_{H}    & M^2_{H}    & 0             \\
m^2_{\Phi} & 0          & M^2_{\Phi}    
\end{array}
\right). 
\label{eq:V''}
\eeq
Here, $m_{H}^2$ and $m_{\Phi}^2$ are $3$-vectors, 
$M_{H}^2$ and $M_{\Phi}^2$ are $3\times3$ matrices. 
$M^2_q $ is the mass term for $H^0_q$.
The form of submatrices $M_H^2$, $M_{\Phi}^2$ 
and sub-vectors $m^2_{H}$, $m^2_{\Phi}$ are fixed 
by the ${\cal T}$-conserving and ${\cal S}$-conserving conditions, 
\begin{eqnarray}
\label{eq:condition}
&&T_{ik}^H M^2_{Hkl} T_{lj}^{H\dag}=M^2_{Hij},~~S_{ik}^\Phi M^2_{\Phi kl}S_{lj}^{\Phi\dag}=M^2_{\Phi ij}, \\\nonumber
&&T_{ik}^H m^2_{Hk}=m^2_{Hi},~~S_{ik}^\Phi m^2_{\Phi k}=m^2_{\Phi i}, 
\end{eqnarray}
where $T^H$ and $S^\Phi$ are generators of subgroups ${\cal T}$ and ${\cal S}$ 
in the triplet representation. 
Similarly, mass matrices for CP-odd and charged scalar mass eigenstates 
are also determined only by the remnant flavor symmetry in each sector. 
We show the most simple example with 
${\cal G}=A_4$, ${\cal T}=Z_3$, and ${\cal S}=Z_2$ in Appendix A. 

Especially, on the base in which $\mathcal{T}$-generator is diagonal, 
subgroup $\mathcal{T}$ restricts the mass terms for scalar bosons 
($H_q^0, H_1^0, H_e^0, H_\phi^0$) 
which interact with the SM particles as follows:
\begin{eqnarray}
\left(\begin{array}{cccc}
M^2_q & m^2_{H \, 1} & 0 & 0 \\
m^2_{H \, 1} & M^2_{H \, 11} & 0 & 0 \\
0 & 0 &  M^2_{H \, 22} & 0 \\
0 & 0 & 0 &  M^2_{H \, 33} 
\end{array}\right). 
\end{eqnarray}
${\cal T}$ completely determines mass matrix for $H_q, H_i$ 
because $\mathcal{G}$-singlet boson $H_q$ also respects 
residual symmetry $\mathcal{T}$. 
Although $H_q^0$, $H_1^0$ and $\Phi_i^0$ may mix each other 
because of non-zero $m^2_{H \,1}$ and $m^2_{\Phi \, i}$, 
the mass matrix of the scalar bosons can be described 
in the model independent way as far as $\Delta V=0$. 

\subsection*{Scalar mass eigenstates in the case of $\Delta V \neq 0$}

Let us consider the case with nonzero $\Delta V$.
We simply assume that the nonzero VEV of $\Phi_i$ is higher than the EW scale, 
and we could write down the ${\cal T}$-conserving and ${\cal T}$-breaking 
effective potential for $H^i$ and $H^q$ at the renormalizable level:
\begin{eqnarray}
V_{eff}&=&V_{\cal T} + V_{\slashed{\cal T}},  \\
V_{\cal T}&=& m^2_q |H_q|^2+(m^2_{q1} H_q^\dagger H_1+h.c.) +m^2_1 |H_1|^2+m^2_2 |H_2|^2+m^2_3 |H_3|^2 +V^{(4)}_T, \nonumber \\
V_{\slashed{\cal T}}&=&m^2_{q2} H_q^\dagger H_2  +m^2_{q3} H_q^\dagger H_3+m^2_{12} H_1^\dagger H_2+m^2_{13} H_1^\dagger H_3+m^2_{23} H_2^\dagger H_3 +h.c.,
\label{eq;T-breaking}
\end{eqnarray}
where $V^{(4)}_T$ is the function which only include 
${\cal T}$-conserving quartic couplings of $H_{q}$ and $H_{i}$. 
$V_{\cal T}$ and $V_{\slashed {\cal T}}$ are the ${\cal T}$-conserving 
and ${\cal T}$-breaking potentials. 
The scalars from $\Phi$ are omitted assuming that 
they decouple below the EW scale 
because of the hierarchy between 
the VEV of $\Phi$ and the VEVs of Higgs doublets. 
$V_{\slashed{\cal T}}$ is generated by $\Delta V$ in Eq. (\ref{potential}). 
We could write quartic couplings in $V_{\slashed {\cal T}}$, 
but the dimensionless couplings are expected to be small, 
because they are generated by the high-dimensional operators in $V$ or 
integrating out the heavy scalars in $\Phi$
\footnote{Our argument in this section would be also reasonable, 
as far as ${\cal T}$-breaking quartic couplings are tiny.}. 

Now, let us consider the stationary conditions for $H_2$ and $H_3$. 
In our setup, they should not develop nonzero VEVs to respect ${\cal T}$. 
The defivatives $\pa_{H_I} V$ ($I=2, \, 3$) 
under the ${\cal T}$-conserving conditions 
depend on the ${\cal T}$-breaking terms as, 
\beq
\pa_{H_I} V_{eff}=m^2_{qI} H_q^\dagger+m^2_{1I} H_1^\dagger. 
\eeq
This leads the equation 
to realize the vacuum alignment where $\langle H_I \rangle=0$ is satisfied: 
\beq
\label{alignment-condition}
m^2_{qI} \sin \beta +m^2_{1I} \cos \beta=0.
\eeq
In general, the stationary conditions for $H_1$ and $H_q$ are independent of 
this condition, so that the fine-tuning would be required, 
if $m^2_{qI}=m^2_{1I}=0$ is not satisfied. 
The quartic couplings in $V_{eff}$ respect the remnant symmetry ${\cal T}$, 
so that $m^2_{qI}$ and $m^2_{1I}$ only contribute to the mass mixing between 
${\cal T}$-charged scalars and ${\cal T}$-trivial scalars. 
We could conclude that ${\cal T}$-charged scalars do not mix 
with ${\cal T}$-trivial scalars, 
unless we admit the tuning in Eq. (\ref{alignment-condition}). 

On the other hand, the mass mixing between $H_2$ and $H_3$ 
generated by $m^2_{23}$ would not be controlled by the vacuum alignment, 
because the both VEVs of $H_2$ and $H_3$ vanish. 
We discuss the ${\cal T}$-breaking effects in Sec. \ref{sec:T-breaking}. 
We also study the ${\cal T}$-breaking terms corresponding to mixing 
between the ${\cal T}$-trivial and the ${\cal T}$-charged scalars, 
in Sec. \ref{section5}.

Note that the scalars in $\Phi_i$ may have the low mass 
compatible with the EW scale, and they may mix with $H_2$ and $H_3$, 
although the mixing given by $\Delta V$ 
should be controlled to realize the vacuum alignment. 
Moreover, $\Phi_i$ dominantly couple with neutrinos, even if the mixing exists. 
Eventually we discuss phenomenology in the limit that the mixing 
with the scalars in $\Phi_i$ is negligible. 

\section{flavor physics}
\label{section4}

We have seen that the FCNCs involving scalars are well controlled, 
if we assume that the partially remnant symmetry ${\cal T}$ is respected 
in the Yukawa couplings. 
Even if ${\cal T}$ is broken in the Higgs potential, 
we could expect that it is possible to discuss the contributions of the FCNCs 
to flavor physics as far as the breaking effect is smaller than the 
${\cal T}$-conserving one. 
As discussed in subsection \ref{sec2-E}, 
the ${\cal T}$-conserving FCNCs are distinguishing, 
so that we could qualitatively analyze their signals in flavor physics. 
In the Sec. \ref{section4-1}, we consider the ${\cal T}$-conserving case 
and then we will see the ${\cal T}$-breaking effect 
including the loop corrections in subsection \ref{sec:T-breaking}. 

\subsection{${\cal T}$-conserving contributions}
\label{section4-1}

First of all, let us discuss the flavor physics in the case that 
${\cal T}$ is conserved in charged leptons and scalar mass matrices.
The Yukawa couplings between scalars and charged leptons are given by 
Eqs. (\ref{eq:Yukawa}) and (\ref{2HDM}). 
$\phi_{e, \, \mu}$ and $H^+_{e, \, \mu}$ are the mass eigenstates in this case. 
The ${\cal T}$-charged scalar masses are also expected 
to be around the EW scale, because the masses are given by $V_H$, 
so we are especially interested in the EW-scale masses of the scalars. 

\subsubsection{${\cal T}$-charged scalar interactions}

Through the exchanging of $\phi_{e, \, \mu}$, 
(flavor-changing) $4$-fermi interactions are effectively generated as 
\begin{eqnarray}
{\cal L}^{(4)}_{\cal T}&=& \frac{1}{v^2 \cos^2 \beta} \Big \{ \frac{|b_3|^2}{m^2_{\phi_e}} (\overline{\tau_R}e_L)(\overline{e_L}\tau_R) + \frac{|b_2|^2}{m^2_{\phi_e}} (\overline{\mu_R}\tau_L)(\overline{\tau_L}\mu_R)
+ \frac{|b_1|^2}{m^2_{\phi_e}} (\overline{e_R}\mu_L)(\overline{\mu_L}e_R) \nonumber  \\
&+&\frac{|c_3|^2}{m^2_{\phi_\mu}} (\overline{\tau_R}\mu_L)(\overline{\mu_L}\tau_R) + \frac{|c_2|^2}{m^2_{\phi_\mu}} (\overline{\mu_R}e_L)(\overline{e_L}\mu_R)
+ \frac{|c_1|^2}{m^2_{\phi_\mu}} (\overline{e_R}\tau_L)(\overline{\tau_L}e_R) \nonumber  \\
&+& \frac{b^*_{2}b_3}{m^2_{\phi_e}} (\overline{\mu_R}\tau_L)(\overline{e_L}\tau_R) + \frac{b^*_1b_2}{m^2_{\phi_e}} (\overline{e_R}\mu_L)(\overline{\tau_L}\mu_R)
+ \frac{b_3^*b_1}{m^2_{\phi_e}} (\overline{\tau_R}e_L)(\overline{\mu_L}e_R) +h.c. \nonumber  \\
&+&\frac{c^*_2c_3}{m^2_{\phi_\mu}} (\overline{\mu_R}e_L)(\overline{\mu_L}\tau_R) + \frac{c_1^*c_2^2}{m^2_{\phi_\mu}} (\overline{e_R}\tau_L)(\overline{e_L}\mu_R)
+ \frac{c^*_3c_1^2}{m^2_{\phi_\mu}}(\overline{\tau_R}\mu_L)(\overline{\tau_L}e_R) +h.c. \Big \}.
\end{eqnarray}
The charged Higgs scalars $H^+_{e, \, \mu}$ also induce flavor violation, 
and it is derived by replacing $\ov{l^i_L}$ with $\ov{\nu_L^k} V_{ki}$. 
The mass difference between $H^+_{e, \, \mu}$ and $\phi_{e, \, \mu}$ 
is strongly constrained by the $\rho$ parameter, 
so that we set $m^2_{\phi_e}=m^2_{H^+_e}$ and $m^2_{\phi_\mu}=m^2_{H^+_\mu}$ 
in our study. 
The lower bound on the charged Higgs mass is given by the direct search 
at the LEP experiment: $m_{H^{\pm}} \gtrsim 80$ GeV \cite{Abbiendi:2013hk}. 
Below, we survey the parameter space above the lower mass. 

\begin{figure}[!t]
{\epsfig{figure=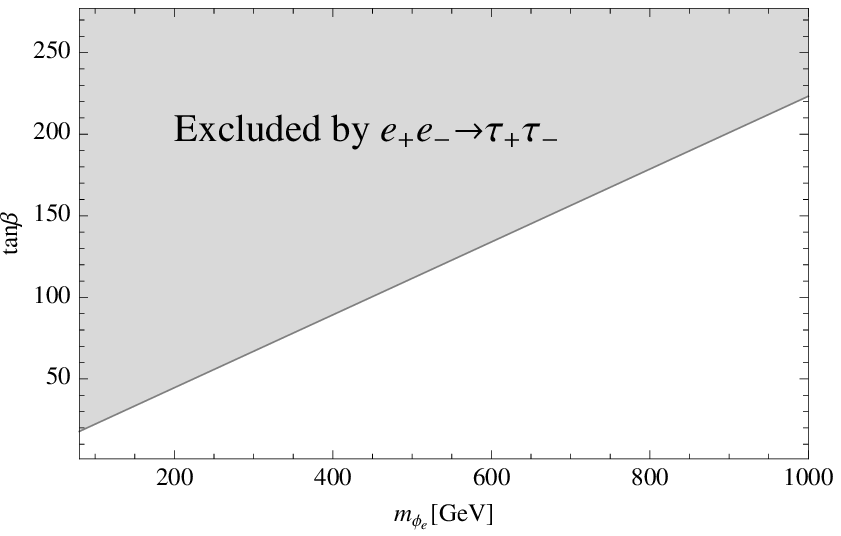,width=0.5\textwidth}}{\epsfig{figure=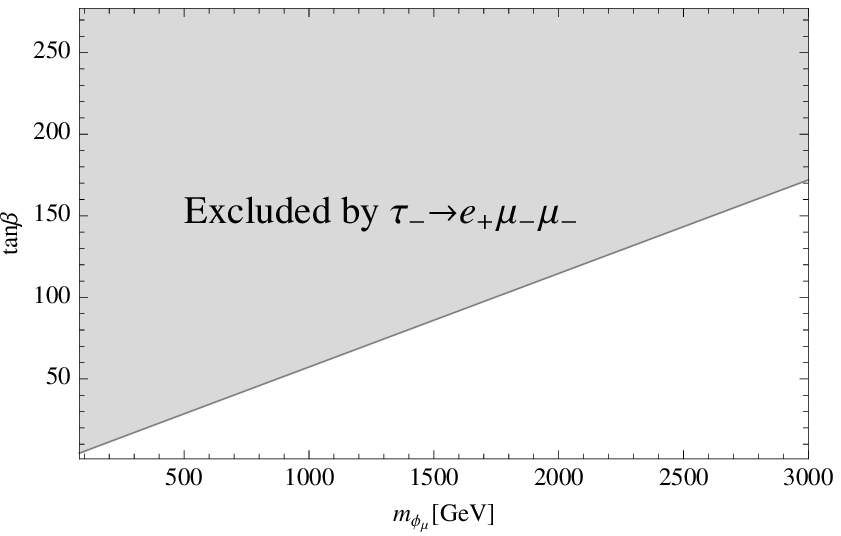,width=0.5\textwidth}}
\vspace{-0.5cm}
\caption{$m_{\phi_{e, \mu}}$ and $\tan \beta$.
The gray region is excluded by $e^+ e^- \to  \tau^+ \tau^-$ at the LEP experiment and the LFV $\tau$ decay,
 $\tau^- \to e^+ \mu^- \mu^-$.}
\label{fig1}
\end{figure}

One of the stringent constraints on the flavor violating couplings is from 
$e^+ e^- \to l^+ l^-$ $(l=e, \, \mu, \, \tau)$ at the LEP \cite{LEPbound}. 
Assuming the relation in Eq. (\ref{A4-relation}) with 
$(m_1, \, m_2, \, m_3)=(m_e, \, m_\mu , \, m_{\tau})$, 
the processes $e^+ e^- \to \mu^+ \mu^-, \,  \tau^+ \tau^-$ 
are enhanced through $\phi_{e \, \mu}$ exchanging in $t$-channel: 
\beq
m_{\phi_e} \gtrsim 0.62 \times \frac{m_{\tau}}{v \cos \beta}~ {\rm TeV},~m_{\phi_\mu} \gtrsim 2.23 \times \frac{m_{\mu}}{v \cos \beta}~ {\rm TeV}.
\eeq
The allowed region is summarized in Fig.\ref{fig1} .
If we consider the model which does not hold the relation 
in Eq. (\ref{A4-relation}), 
$m_{\phi_e}$ may face the stronger bound from $e^+ e^- \to \mu^+ \mu^-$, 
but the bound from the flavor violating $\tau$ decay is stronger 
than the LEP bound, as we discuss below. 

Flavor violating decays of $\tau$ and $\mu$ have been 
well investigated in the experiments, 
and the constraints are summarized in 
Refs. \cite{Hayasaka:2010np, Bellgardt:1987du}. 
In our models, ${\cal T}$-charge should be conserved, so that 
the final states from $\tau$ and $\mu$ decays should be ${\cal T}$-charged states. 
That is, the possible decay patterns of $\tau$ are only 
\beq
\tau^- \to \mu^+ e^- e^-,~ e^+ \mu^- \mu^-.
\eeq
$\mu \to 3e$, which is strongly constrained 
by the experiments \cite{Bellgardt:1987du}, can be forbidden. 
Assuming the relation in Eq. (\ref{A4-relation}), 
the strong bound on the flavor violating decays is 
obtained from mainly $\tau^- \to  e^+ \mu^- \mu^-$ as 
Br($\tau^- \to e^+ \mu^- \mu^-$) $< 1.7 \times 10^{-8}$ \cite{Hayasaka:2010np} with 
\beq
{\rm Br}(\tau^- \to e^+ \mu^- \mu^-)=\frac{m^5_\tau}{3 (8 \pi)^3 \Gamma_\tau} \left | \frac{ m_\tau m_\mu}{m^2_{\phi_\mu} (v\cos \beta)^2}  \right |^2.
\eeq
The allowed region is summarized in Fig.\ref{fig1} \footnote{The lepton flavor violating (LFV) $\tau$ decays induced by tree-level FCNCs involving extra scalars have been investigated in a generic two-Higgs-doublet model \cite{LFV-2HDM}. }.

In addition, the charged Higgs exchanging processes 
may also contribute to the $\tau$ and $\mu$ decay. 
The chirality of the charged lepton in the decays 
is different from the one in the SM, 
so that the correction my be quite small. 
Assuming the charged lepton in the final state is massless, 
the deviation of the leptonic decay is evaluated as, 
\beq
\Delta{\rm Br}(l_i \to l_j \ov{\nu_k} \nu_n) \simeq \frac{1}{32 G^2_F} \sum_{a=e, \mu} \frac{|Y_a^{ni} Y_a^{* \, kj}|^2}{m^4_{H^+_a}}. 
\eeq
Assuming the relation in the Eq. (\ref{A4-relation}) 
and $m_{H^+_a}=m_{\phi_a}$ for the EWPOs, 
we find that the modified branching ratio of muonic $\tau$ decay is at most, 
\beq
{\rm Br}(\tau  \to \mu \ov{\nu} \nu) \simeq (1-1.07 \times 10^{-3}) \times {\rm Br}(\tau  \to \mu \ov{\nu} \nu)_{\rm SM},
\eeq
including the contribution of $\mu$ mass. 
${\rm Br}(\tau  \to \mu \ov{\nu} \nu)_{\rm SM}$ is the SM prediction and
this modified value is within the error of the current experimental measurement 
of the $\tau$ decay \cite{PDG}. 
The contribution of charged Higgs carrying also ${\cal T}$-charges to the other decay 
such as $\mu \to e \nu \nu$ is strongly suppressed by the Yukawa couplings. 

Including one-loop corrections involving the extra scalars, 
the $Z$ and $W$ couplings would be slightly deviated from the SM ones:
\begin{equation}
{\cal L}_{{\rm EW}}= g_Z Z^{\rho} \{  (q_L +\Delta q^i_L) \ov{l^i_L} \gamma_\rho l^i_L+(q_R+\Delta q^i_R) \ov{l^i_R} \gamma_\rho l^i_R+  (q^{\nu}_L +\Delta q^{\nu \, ij}_L) \ov{\nu^i_L} \gamma_\rho \nu^j_L \}, 
\end{equation}
where 
$(q_L, \, q_R, \, q^{\nu}_L)=(-1/2+\sin^2 \theta_W, \, \sin^2 \theta_W, \, 1/2)$ 
are defined. $\sin \theta_W$ is the Weinberg angle, and $g_Z$ is the gauge coupling for $Z$-boson interaction.
According to the one-loop diagrams involving $\phi_{e \, \mu}$, 
they are estimated as follows: 
\begin{eqnarray}
\Delta q^{i}_L &=& \sum_{a=e,\mu} \sum^3_{k=1} \frac{|Y^{ik}_a|^2}{16 \pi^2} \frac{M^2_Z}{m^2_{\phi_a}} \left (- \frac{1}{36}- \frac{1}{3} \sin^2 \theta_W \right ),  \\
\Delta q^{i}_R &=&\sum_{a=e,\mu} \sum^3_{k=1} \frac{|Y^{ki}_a|^2}{16 \pi^2} \frac{M^2_Z}{m^2_{\phi_a}} \left (\frac{7}{36}- \frac{1}{3} \sin^2 \theta_W \right ), \\
\Delta q^{\nu \, ij}_L &=& \sum_{a=e,\mu} \sum^3_{k,l,n=1}V^{*\,ki} V^{ nj} \frac{Y^{kl}_a Y^{nl}_a}{16 \pi^2} \frac{M^2_Z}{m^2_{\phi_a}} \left (\frac{1}{36}- \frac{7}{18} \sin^2 \theta_W \right ),
\end{eqnarray}
where $m_i \ll M^2_Z \ll  m_{\phi_a}$ is assumed. 
In the model which satisfies the relation in Eq. (\ref{A4-relation}), 
$\Delta q^{e, \mu}_L$, $\Delta q^{\tau}_R$, and $\Delta q^{\nu \, ij}_L$ 
might be sizable. 
The constraints from $e^+e^- \to \tau^+ \tau^-$ and $\tau^- \to e^+ \mu^- \mu^-$ 
give the maximal sizes of the deviations: 
\begin{eqnarray}
\Delta q^e_{L}&\approx&2.87 \times 10^{-5}, ~\Delta q^\mu_{L}\approx-1.89 \times 10^{-6},~\Delta q^\tau_{R}\approx3.21 \times 10^{-5},  \\
\Delta q^{\nu \, ij}_{L}&\approx&-1.7 \times 10^{-5} \times V^{i3}  V^{* \,j3}.  
\end{eqnarray}
They are too tiny to compare with the current experimental results. 
In fact, the maximal values are within the error of the measurements 
of $Z$ boson \cite{Pich:2013lsa}. 

\subsubsection{${\cal T}$-trivial scalar interactions}

$H_q$ and $H_1$ are not charged under ${\cal T}$, and they develop nonzero VEVs. 
Their Yukawa couplings with SM fermions are flavor-diagonal in the mass base, 
under the ${\cal T}$-conserving assumption. 
Then, we could conclude that $H_q$ and $H_1$ induce so-call minimal flavor 
violation, and evade the stringent constraints from flavor physics. 
This type of scalars with the interactions in Eq. (\ref{2HDM}) 
have been well investigated so far, motivated by, for instance, 
the deviation of muon anomalous magnetic moment \cite{leptophilic2HDM}. 

The lower bound on the charged Higgs mass 
comes from the direct search for charged Higgs at the LHC 
and it is given by top mass to evade the exotic top decay: 
$m_{H^{\pm}} \gtrsim 172$ GeV \cite{chargedLHC}. 
The pseudo scalar mass should be close to the charged Higgs mass 
to avoid too large deviation of the $\rho$ parameter. 
Flavor changing processes in $B$ physics may constrain our ${\cal T}$-trivial scalars. 
For instance, $B \to X_s \gamma$ process gives the lower bound on $\tan \beta$: 
$\tan \beta \gtrsim 1$ \cite{Hermann:2012fc}. 
Br$(B^- \to \tau^- \nu)$ is also slightly deviated from the SM prediction 
according to the charged Higgs exchanging, but it is less than $1 \%$ 
in the region with $m_{H^\pm} \geq 172$ GeV. 
As pointed out in Ref. \cite{leptophilic2HDM}, 
the discrepancy of muon anomalous magnetic moment 
may be explained if $\tan \beta$ is quite large and pseudo scalar is rather small. 
Let us also survey the parameter region in the next section. 

%%------------------------------------------------------------------------------
%------------------------------------------------------------------------------

\subsection{${\cal T}$-breaking contributions}
\label{sec:T-breaking}

Next, we investigate the contributions of ${\cal T}$-breaking terms 
to flavor physics. 
If we can assume that the remnant symmetry ${\cal T}$ is respected 
in Higgs potential after the symmetry breaking, 
we could expect that only ${\cal T}$-symmetric terms 
are relevant to flavor violating processes. 
However, as we have seen in Sec. \ref{section3}, 
$\Delta V$ may be allowed 
even if we assume that the vacuum alignment respects ${\cal T}$ and ${\cal S}$. 
After the symmetry breaking, $\Delta V$ induces ${\cal T}$-breaking terms, 
such as $m^2_{23}$ in Eq. (\ref{eq;T-breaking}), 
in scalar mass matrices because of nonzero $\langle \Phi \rangle$, 
so that we may have to control $\Delta V$ to evade 
the stringent constraints from flavor physics. 

${\cal T}$-breaking terms would appear in scalar mass matrices, 
according to $\Delta V$ in Eq. (\ref{potential}), 
\begin{eqnarray}
{\cal L}_{\cal T} = -\frac{1}{2} (\delta m^2_H)_{ab} H^0_a H^0_b -\frac{1}{2} (\delta m^2_A)_{ab} A_a A_b - (\delta m^2_{H^+})_{ab} H_a^+ H_b^-. 
\label{deltam}
\end{eqnarray}
$A_{a}$ and $H_a^+$ denote the two kinds of scalars: 
$\{A_{a}\}=\{A_e, A_\mu \}$ and $\{H^{\pm}_{a}\}=\{ H^{\pm}_e, H^{\pm}_\mu \}$.
There are two CP-even scalars 
and they mix each other in general according to $(\delta m^2_H)_{ab}$, 
where $\{H_a^0\}=\{ H^0_{e}, H^0_{\mu}\}$ is defined. 

Besides, there may be mixings between ${\cal T}$-trivial and ${\cal T}$-charged scalars,
such as $(\delta m^2_H)_{a1} H^0_a H^0_{S \, 1}$, although they require the fine-tuning 
against the parameters in Higgs potential, as discussed in Eq. (\ref{alignment-condition}).
The mixings relate to the vacuum alignment, so we analyze the ${\cal T}$-breaking terms
including the study about the deviation of the vacuum alignment, in Sec. \ref{section5}.

In general, $\Phi$ also predicts extra scalars, 
and should be involved in the scalar mass matrices. 
However, $\Phi$ mainly couples with neutrinos, 
so the constraint on $\Phi^i$ is rather weak. 
Simply we assume that the scalars from $\Phi^i$ gain heavy masses 
according to nonzero VEVs of $\Phi^i$ in $V_{\Phi}$, 
and decouple around the EW scale. 

We could expect that this assumption 
leads the approximate ${\cal T}$-conserving situation with 
\beq
(\delta m^2_H)_{ab}, \, (\delta m^2_A)_{ab}, \, (\delta m^2_{H^+})_{ab} \ll m^2_{\phi_{e,\mu}}, 
\eeq
and discuss the bound on the breaking effects from the experiments. 
The relevant constraints are from $l \to l' \gamma$ processes \cite{LFV-2HDM-2, Chang:1993kw}. 
Especially, main contribution of the ${\cal T}$-breaking terms 
would appear in $\mu \to e \gamma$. 

\subsubsection{constraint from the $\mu \to e \gamma$ process}
\label{mutoegamma}

The $\mu \to e \gamma$ process has been well investigated 
in 2HDMs \cite{LFV-2HDM-2, Chang:1993kw}. 
The MEG experiment released the upper bound 
on the branching ratio of the flavor changing process: 
Br$(\mu \to e \gamma) < 5.7 \times 10^{-13}$ \cite{Adam:2013mnn}. 
It would be updated up to $6 \times 10^{-14}$ in the future \cite{Baldini:2013ke}. 

Our dominant contribution to the $\mu \to e \gamma$ process 
is from the one-loop correction involving the scalars of $\phi_e$, 
because $\phi_e$ has large $(e, \, \tau)$ and $(\tau,\, \mu)$ elements 
of the flavor-violating Yukawa couplings. 
If the CP-even scalar and CP-odd scalar masses of $\phi_e$ are different, 
the $\mu \to e \gamma$ process is easily enhanced. 
The operator to induce the LFV process is estimated as follows at the one-loop level: 
\begin{eqnarray}
{\cal L}_{\mu \to e \gamma}&=&e C_7 \overline{e_L} \sigma_{\mu \nu} \mu_RF^{\mu \nu}, \\
C_7&=&\frac{ m_{\tau}Y^{e \tau }_e Y^{ \tau \mu }_e}{64\pi^2}  \left \{ \frac{U^h_{e \alpha} U^h_{e \alpha}}{m^2_{h_\alpha}}  F  \left( m^2_{h_\alpha}/m^2_\tau \right ) - \frac{U^A_{e \alpha} U^A_{e \alpha} }{m^2_{A_\alpha}} F \left( m^2_{A_\alpha}/m^2_\tau \right ) \right \},
\end{eqnarray}
where $e$ is the electric charge and $F(x)$ is defined as
\beq
F (x)= \ln ( x ) -\frac{3}{2}.
\eeq

\begin{figure}[!t]
{\epsfig{figure=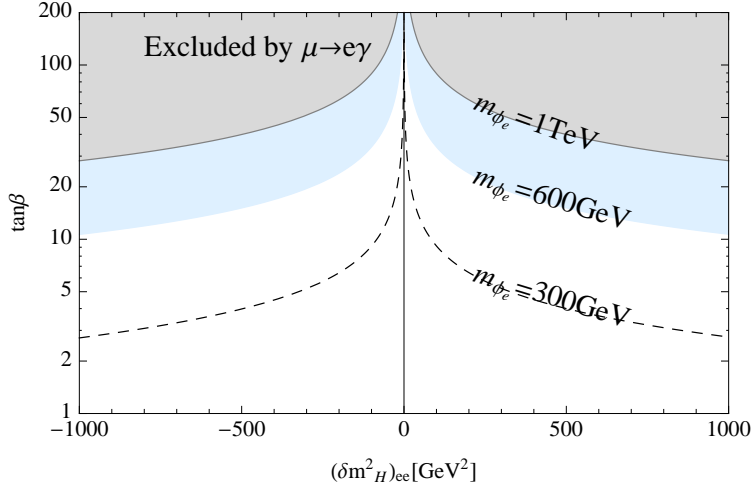,width=0.6\textwidth}}
\vspace{-0.5cm}
\caption{$(\delta m^2_H)_{ee}$ (GeV) and $\tan \beta$. The gray (light-blue) is excluded by $\mu \to e \gamma$ at $m_{\phi_{e}}=1000$ ($600$) GeV. The dashed line is the upper bound with $m_{\phi_{e}}=300$ GeV.
}
\label{fig2}
\end{figure}

$U^{h,A}_{ij}$ are the diagonalizing matrices for the mass matrices of CP-even and -odd scalars;
\begin{eqnarray} 
((U^h)_{a \alpha} m^2_{h_\alpha} (U^h)_{b\alpha})&=&\begin{pmatrix} m^2_{\phi_e} + (\delta m_H^2)_{ee} & (\delta m_H^2)_{\mu e} \\  (\delta m_H^2)_{\mu e} & m^2_{\phi_\mu}+  (\delta m_H^2)_{\mu\mu} \end{pmatrix}, 
\label{Uh}
\end{eqnarray}
\begin{eqnarray}
((U^A)_{a \alpha} m^2_{A_\alpha} (U^A)_{b \alpha})&=&\begin{pmatrix} m^2_{\phi_e}+(\delta m_A^2)_{ee} & (\delta m_A^2)_{\mu e} \\ (\delta m_A^2)_{\mu e} & m^2_{\phi_\mu}+ (\delta m_A^2)_{\mu\mu} \end{pmatrix}. 
\label{UA}
\end{eqnarray}
Fig. \ref{fig2} shows the excluded region by the current upper bound on 
$\mu \to e \gamma$ process, in the case that only $(\delta m^2_H)_{ee}$ is nonzero. 
As we see, the mass difference between the CP-even and CP-odd scalars 
is severely constrained by the flavor-changing process. 
If $\phi_e$ is below $300$ GeV, $\tan \beta$ should be smaller than about $2$. 
This is quite strong, 
compared to the bound from $e^+e^- \to \tau^+ \tau^-$ in Fig. \ref{fig1}. 
We may require large $\tan \beta$, for instance, 
to enhance the muon anomalous magnetic moment \cite{leptophilic2HDM}, 
but the ratio of the squared mass difference to $m^2_{\phi_e}$, $(\delta m^2_H)_{ee}/m^2_{\phi_e}$, should be much smaller than $O(10^{-2})$. 

%------------------------------------------------------------------------------
%------------------------------------------------------------------------------

\subsubsection{constraint from the $\tau \to e \gamma$ process}

\begin{figure}[!t]
{\epsfig{figure=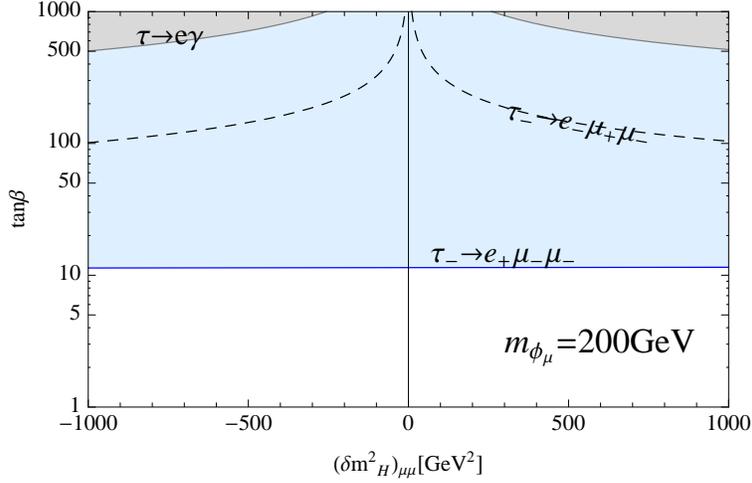,width=0.6\textwidth}}
\vspace{-0.5cm}
\caption{$(\delta m^2_H)_{\mu \mu}$ (GeV) and $\tan \beta$. The gray (light-blue) is excluded by $\tau \to e \gamma$ at $m_{\phi_{\mu}}=200$ GeV. The region above the dashed line is the upper bound from $\tau^- \to e^- \mu^+ \mu^-$
and the blue region is the excluded one by $\tau^- \to e^+ \mu^- \mu^-$.
}
\label{fig3}
\end{figure}

The scalar of $\phi_\mu$ contribute to the $\tau \to e \gamma$ process, 
according to the mass difference between the CP-even and CP-odd scalar. 
The operator for the LFV process is given by 
\begin{eqnarray}
{\cal L}_{\tau \to e \gamma}&=&e C^{\tau}_7 \overline{e_L} \sigma_{\mu \nu} \tau_RF^{\mu \nu}, \\
C^{\tau}_7&=&\frac{ m_{\mu}Y^{e \mu }_\mu Y^{  \mu\tau }_\mu}{64\pi^2}  \left \{ \frac{U^h_{\mu \alpha} U^h_{\mu \alpha}}{m^2_{h_\alpha}}  \left( F \left( m^2_{h_\alpha}/m^2_\mu \right )  + \frac{m_\tau}{6m_\mu}\right )- \frac{U^A_{\mu \alpha} U^A_{\mu \alpha} }{m^2_{A_\alpha}} \left( F \left( m^2_{A_\alpha}/m^2_\mu \right )  + \frac{m_\tau}{6m_\mu} \right ) \right \}. \nonumber \\
\end{eqnarray}
The current upper bound on the $\tau \to e \gamma$ process is 
$1.1 \times 10^{-7}$ \cite{Aubert:2009ag}, 
and it is rather weak compared with the one from the $\mu \to e \gamma$ process. 
In Fig. \ref{fig3}, the regions excluded by 
$\tau \to e \gamma$, $\tau^- \to e^- \mu^+ \mu^-$, and $\tau^- \to e^+ \mu^- \mu^-$ 
are summarized. We can conclude that the $\tau^- \to e^+ \mu^- \mu^-$ constraint 
is the most important to $\phi_\mu$ in our model, even if we include the ${\cal T}$-breaking terms.
%------------------------------------------------------------------------------
%------------------------------------------------------------------------------

\subsubsection{muon anomalous magnetic moment  $(g-2)_{\mu}$ }

In our model, the ${\cal T}$ breaking term 
which allows the mass mixing between $\phi_e$ and $\phi_\mu$ 
enhances the muon anomalous magnetic moment and electron, muon EDMs. 

It is well-known that there is a $3.1$ $\sigma$ deviation from the SM prediction 
in the muon anomalous magnetic moment $(g-2)_{\mu}$ experimental result \cite{PDG}. 
In Ref. \cite{leptophilic2HDM}, very light pseudo-scalar is introduced 
to achieve the anomaly in the leptophilic 2HDM. 
In our model, we could find new contributions to $(g-2)_{\mu}$ 
according to the tree-level FCNCs \cite{LFV-2HDM-2}, 
\beq
\label{muong-2}
\Delta a_\mu=\frac{ m_\mu m_\tau Y^{ \tau \mu}_e Y^{  \mu \tau }_\mu }{(4 \pi)^2} \left \{  \frac{U^{h}_{e \alpha} U^h_{\mu \alpha} }{m^2_{h_\alpha}} F \left( m^2_{h_\alpha}/m^2_\tau \right )  - \frac{ U^A_{e \alpha} U^A_{\mu \alpha}}{m^2_{A_\alpha}}  F \left(m^2_{A_\alpha}/m^2_\tau \right )  \right \}.
\eeq
Unfortunately, the enhancement of $\Delta a_{\mu}$ is tiny 
as long as the ${\cal T}$-breaking terms are small, 
because of the stringent constraint from 
the $\mu \to e \gamma$ and the $\tau^- \to e^+ \mu^- \mu^-$ processes. 
Setting $(\delta m^2_{H,A})_{ee}=(\delta m^2_{H,A})_{\mu}=(\delta m^2_{A})_{e\mu}=0$, 
$\Delta a_\mu$ is at most $O(10^{-2}) \times 10^{-9}$, 
which is much below the experimental result. 
One possible way to enhance $\Delta a_\mu$ would be large $\tan \beta$ 
and light ${\cal T}$-trivial scalar scenario, 
as pointed out in Ref. \cite{leptophilic2HDM}. 

In addition, the loop corrections involving extra scalars 
with ${\cal T}$-breaking terms deviate the mass base of the charged leptons. 
As long as $\tan \beta$ is rather small, the deviation would be tiny 
but the large $\tan \beta$ scenario may be also fascinating 
because of the discrepancy of $(g-2)_\mu$. 
Furthermore, nonzero $\theta_{13}$ is confirmed at the experiments \cite{theta13-1,theta13-2,theta13-3,theta13-4,theta13-5}, 
so it is important to discuss the contribution of the ${\cal T}$-breaking terms 
to the PMNS matrix.
In the next section, we investigate the mass mixing from the one-loop correction, 
and discuss the contribution to $\theta_{13}$ and the flavor changing processes. 

\begin{figure}[!t]
{\epsfig{figure=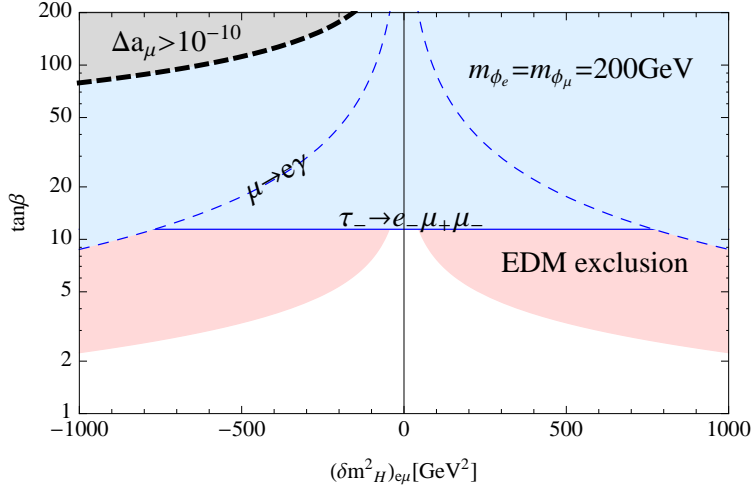,width=0.6\textwidth}}
\vspace{-0.5cm}
\caption{$(\delta m^2_H)_{e \mu}$ (GeV) and $\tan \beta$ with $m_{\phi_{e}}=m_{\phi_{\mu}}=200$ GeV. 
}
\label{fig4}
\end{figure}

When Yukawa couplings in Eq. (\ref{muong-2}) are complex, 
contributions to electric dipole moment occur from their imaginary parts. 
The electron and muon EDMs are given as follows: 
\begin{eqnarray}
d_e&=&\frac{e }{32\pi^2} {\rm Im}(Y_e^{e\tau}Y_\mu^{\tau e}) \left\{U_{\mu \alpha}^hU_{e \alpha}^h\frac{m_\tau}{m_{h \alpha}^2}F\left(m_{h \alpha}^2/m_\tau^2\right)
-U_{\mu \alpha}^AU_{e \alpha}^A\frac{m_\tau}{m_{A \alpha}^2}F \left(m_{A \alpha}^2/m_\tau^2\right)\right\}, \nonumber \\ && \\
d_\mu&=&\frac{e }{32\pi^2} {\rm Im}(Y_\mu^{\mu\tau}Y_e^{\tau \mu}) \left\{U_{\mu \alpha}^hU_{e \alpha}^h\frac{m_\tau}{m_{h \alpha}^2}F\left(m_{h \alpha}^2/m_\tau^2\right)
-U_{\mu \alpha}^AU_{e \alpha}^A\frac{m_\tau}{m_{A \alpha}^2}F\left(m_{A \alpha}^2/m_\tau^2\right)\right\}. \nonumber \\ 
\end{eqnarray}
The current upper bounds on the electron and muon EDMs are 
$|d_e|<8.7 \times 10^{-29} [e{\rm cm}]$ \cite{EDM} and 
$|d_\mu|<1.8 \times 10^{-19} [e{\rm cm}]$ \cite{Bennett:2008dy}, respectively. 
Fig. \ref{fig4} shows the allowed region 
in the case with $m_{\phi_{e}}=m_{\phi_{\mu}}=200$ GeV. 
The imaginary parts of ${\rm Im}(Y_e^{ij}Y_\mu^{kl})$ are assumed 
to be given by Eqs. (\ref{A4-relation}) and (\ref{eq:Yukawa}). 
The red region in Fig. \ref{fig4} shows the excluded region by the current experimental bound on 
the electron EDM. Depending on the phase of the Yukawa couplings, it is the most stringent one among 
the relevant constraints in our model. 

%------------------------------------------------------------------------------
%------------------------------------------------------------------------------

\subsection{short summary }
\begin{figure}[!t]
{\epsfig{figure=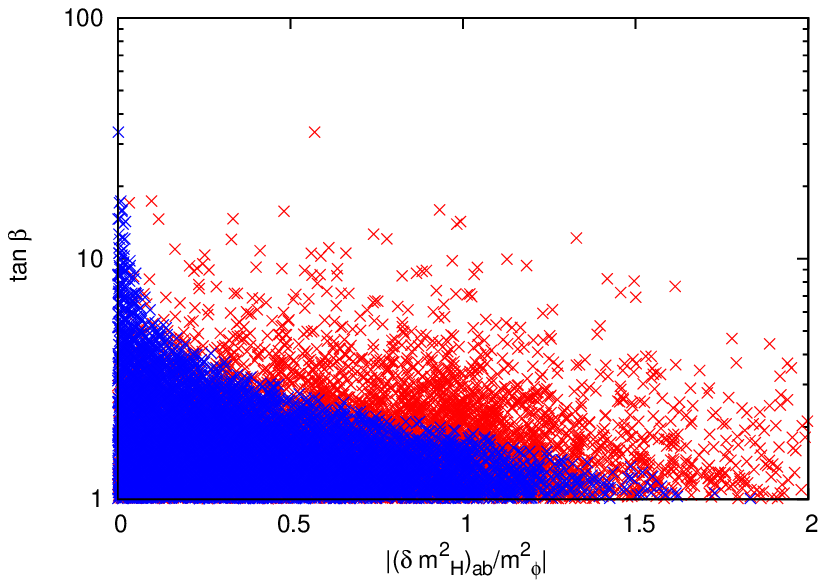,width=0.5\textwidth}}{\epsfig{figure=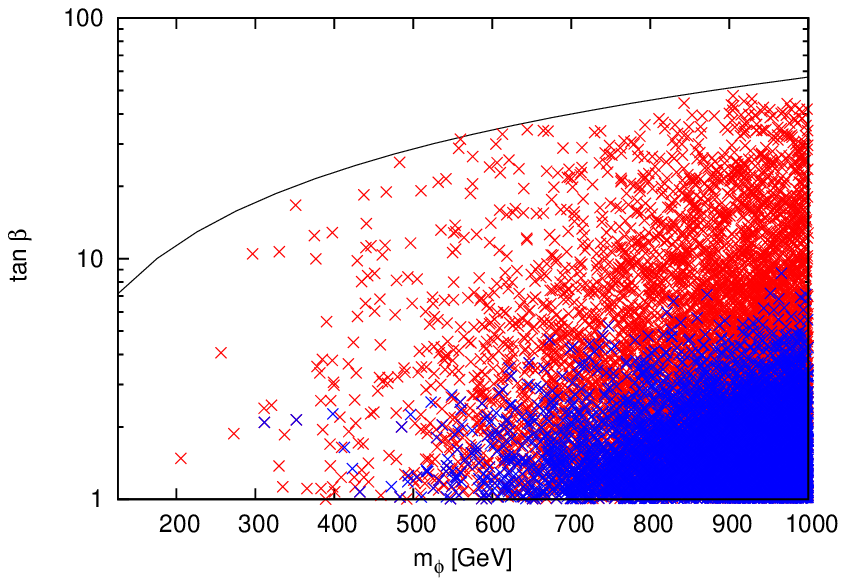,width=0.5\textwidth}}
\vspace{-0.5cm}
\caption{ $|(\delta m^2_H)_{ab}/m^2_{\phi}|$ and $\tan \beta$ (left), and $m_\phi$ (GeV) and $\tan \beta$ (right). $m_{A_e}$ and $m_{A_\mu}$ are fixed at $m_\phi$, and $m_{\phi}$ is within 
$126$ GeV and $1$ TeV range. In the left figure,  blue points figure out $|(\delta m^2_H)_{\mu e}/m^2_{\phi}|$ and $|(\delta m^2_H)_{ee}/m^2_{\phi}|$, and red points are $(\delta m^2_H)_{\mu \mu}/m^2_{\phi}$. 
In the right figures, $|(\delta m^2_H)_{ab}/m^2_{\phi}|$ is larger than $0.01$. The blue (red) points are the allowed ones
for the experimental bounds with (without) the upper limit of the electron EDM. The black line is the upper bound 
in the case with $|(\delta m^2_H)_{ab}|=0$. 
}
\label{fig-summary}
\end{figure}

Let us summarize the results in this section. 
We investigate the experimental bounds from flavor physics. 
In the ${\cal T}$-conserving case, the lepton flavor violating decay, 
$\tau^- \to e^+ \mu^- \mu^- $, gives the stringent constraint. 
If we include the ${\cal T}$-breaking terms in the scalar mass matrices, 
$\mu \to e \gamma$ and the electron EDM are relevant to our model. 
We summarize the allowed points in Fig. \ref{fig-summary}. 
$m_{A_e}$ and $m_{A_\mu}$ are set to be equal to $m_{\phi}$ 
and they are within $126$ GeV and $1$ TeV range. 
The left figure of Fig. \ref{fig-summary} shows the allowed regions 
for the ${\cal T}$-breaking terms: 
the blue points correspond to $|(\delta m^2_H)_{\mu e}/m^2_{\phi}|$ and $|(\delta m^2_H)_{ee}/m^2_{\phi}|$ respectively,
and the red points figure out $|(\delta m^2_H)_{\mu \mu}/m^2_{\phi}|$.
As discussed in this section, $(\delta m^2_H)_{ee}$ and $(\delta m^2_H)_{\mu e}$
are strongly constrained by $\mu \to e \gamma$ and the electron EDM, while the bound on
$(\delta m^2_H)_{\mu \mu}$ is relatively weak. If we take $\tan \beta$ to be larger than $10$, 
the ${\cal T}$ breaking terms should be less than $O(0.1)$ compared with the ${\cal T}$-conserving parts.
In other words, the upper bound on $\tan \beta$ is less than $10$, if $|(\delta m^2_H)_{\mu e}/m^2_{\phi}|$
is larger than $O(0.1)$.

In the right figure of Fig. \ref{fig-summary}, we see the allowed region for $m_{\phi}$ and $\tan \beta$.
$|(\delta m^2_H)_{ab}/m^2_{\phi}|$ is larger than $0.01$. The black line corresponds to the upper limit
with the vanishing $(\delta m^2_H)_{ab}$.
The red (blue) points are the allowed ones without (with) the constraint from the electron EDM.
$Y_e^{e\tau}Y_\mu^{\tau e}$ is assumed to be pure imaginary on the blue points.
As we see, the light $m_{\phi}$ is disfavored by the $\mu \to e \gamma$ process,
while the bound from  $\tau^- \to e^+ \mu^- \mu^- $ is more important in the heavy $m_{\phi}$ region.
If $Y_e^{e\tau}Y_\mu^{\tau e}$ includes imaginary parts, the electron EDM may become
more relevant to our model, and give the severe constraint as in Fig. \ref{fig-summary}.

%------------------------------------------------------------------------------
%------------------------------------------------------------------------------

\section{ Mass mixing induced by ${\cal T}$-breaking terms}
\label{section5}

Off-diagonal elements of Dirac mass matrix for charged leptons are generated by 
loop diagrams including interactions coming from $\Delta V$,
where the mixing between scalar bosons 
carrying different ${\cal T}$-charges occurs as shown in Eq. (\ref{deltam}). 
Including corrections for Dirac mass matrix of charged leptons  ($M_l$), 
let us redefine the mass matrix 
\beq
M_l=\left(
\begin{array}{ccc}
m_e               & \epsilon_{e\mu}    & \epsilon_{e\tau}   \\
\epsilon_{\mu e}  & m_\mu              & \epsilon_{\mu\tau} \\
\epsilon_{\tau e} & \epsilon_{\tau\mu} & m_\tau             
\end{array}\right).
\eeq
%------------------------------------------------------------------------------
\begin{figure}[h]
 \begin{center}
  \epsfig{file=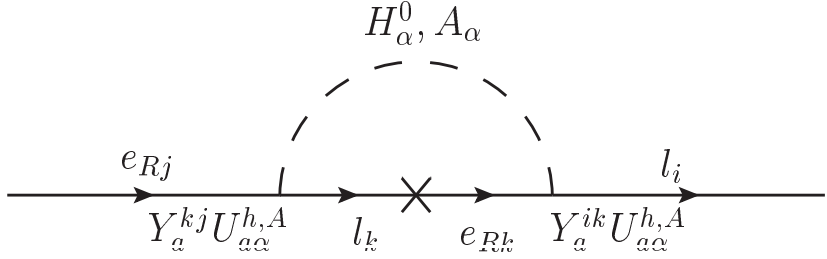,width=0.7\hsize}
   \caption{1-loop diagram which gives $\epsilon_{ij}$. 
            $H^0_a, A_a$ in this figure is scalar mass eigenstates 
            in ${\cal T}$-breaking case.}
  \label{fig:tbreaking}
 \end{center}
\end{figure}
%------------------------------------------------------------------------------
In order to derive explicit representation for $\epsilon_{ij}$, 
we consider 1-loop processes as shown in Fig. \ref{fig:tbreaking}. 
The off-diagonal elements of $M_l$ are estimated as, 
\begin{eqnarray}
\label{eq;neutrino-mixing}
\epsilon_{ij}=\sum_{a,\alpha,\beta}\frac{Y_a^{ik}m_kY_b^{kj}}{32\pi^2}\left\{U_{a \alpha}^hU_{b \alpha}^h\ln{\frac{m_{h_\alpha}^2}{\Lambda^2}}-U_{a \alpha}^AU_{b \alpha}^A\ln{\frac{m_{A_\alpha}^2}{\Lambda^2}}\right\}, 
\end{eqnarray}
where $i, \, j, \, k$ indices represent charged leptons, and
$a, \, b=e, \, \mu$ are defined. 
$\Lambda$ is some scale, but $\epsilon_{ij}$ do not explicitly depend on $\Lambda$. 
We assume $m_k\ll m_{h_\alpha}$, $m_{A_\alpha}$.  

These loop corrections change the mass base slightly, 
and would contribute to the physical observables such as neutrino mixing angles. 
Here, we investigate how large $\theta_{13}$ can be 
according to the radiative correction and discuss the correlation 
between the neutrino mixing and the predicted flavor changing process. 

On the other hand, we may find extra FCNCs generated by the radiative corrections. 
For instance, the Yukawa couplings of the ${\cal T}$-trivial scalars 
are flavor-diagonal at the tree level, 
but nonzero off-diagonal elements appear at the one-loop level, 
according to the radiative correction to the Yukawa couplings 
involving the ${\cal T}$-trivial scalars. 
The one-loop FCNCs can be described as, 
\begin{eqnarray}
{\cal L}^{(1)}_{FCNC}&=& -y^{1}_{ij} H^0_{S \, 1} \ov{l}_i e_{R \, j} -y^{2}_{ij} H^0_{S \, 2}  \ov{l}_i e_{R \, j},  \\
y^1_{ij}&=&\sum_{a, \alpha, \beta}\frac{Y_\alpha^{ik}m_kY_\beta^{kj}}{16\pi^2 \sqrt{2}} \left ( \cos \alpha \frac{\pa}{ \pa \langle H_q \rangle} +\sin \alpha \frac{\pa}{ \pa \langle H_1 \rangle} \right ) \left\{U_{\alpha a}^hU_{\beta a}^h\ln{\frac{m_{h_a}^2}{\Lambda^2}}-U_{\alpha a}^AU_{\beta a}^A\ln{\frac{m_{A_a}^2}{\Lambda^2}}\right\}, \nonumber  \\
&& \\
y^2_{ij}&=&\sum_{a, \alpha, \beta}\frac{Y_\alpha^{ik}m_kY_\beta^{kj}}{16\pi^2 \sqrt{2}} \left ( \cos \alpha \frac{\pa}{ \pa \langle H_1 \rangle} -\sin \alpha \frac{\pa}{ \pa \langle H_q \rangle} \right ) \left\{U_{\alpha a}^hU_{\beta a}^h\ln{\frac{m_{h_a}^2}{\Lambda^2}}-U_{\alpha a}^AU_{\beta a}^A\ln{\frac{m_{A_a}^2}{\Lambda^2}}\right\}. \nonumber  \\
&&
\end{eqnarray} 

\begin{figure}[h]
 \begin{tabular}{cc}
  \begin{minipage}{0.5\hsize}
   \begin{center}
    \epsfig{file=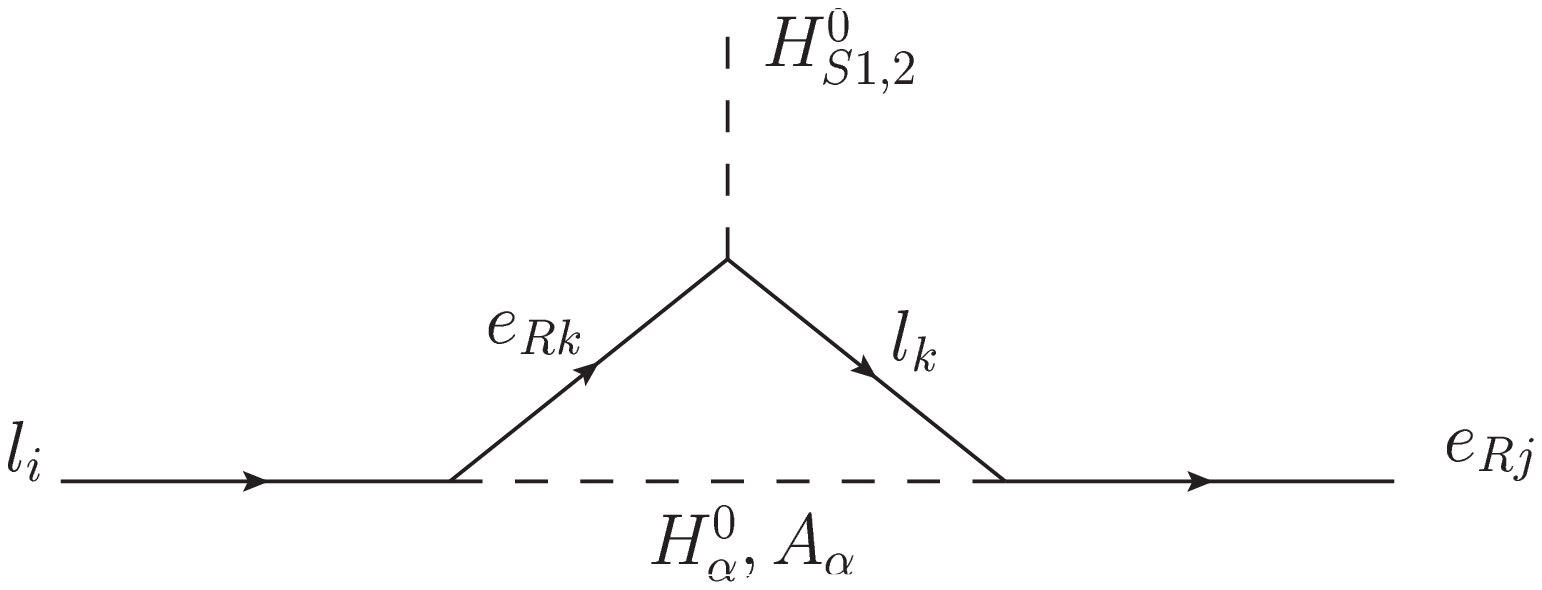,width=\hsize}
   \end{center}
  \end{minipage}
  \begin{minipage}{0.5\hsize}
   \begin{center}
    \epsfig{file=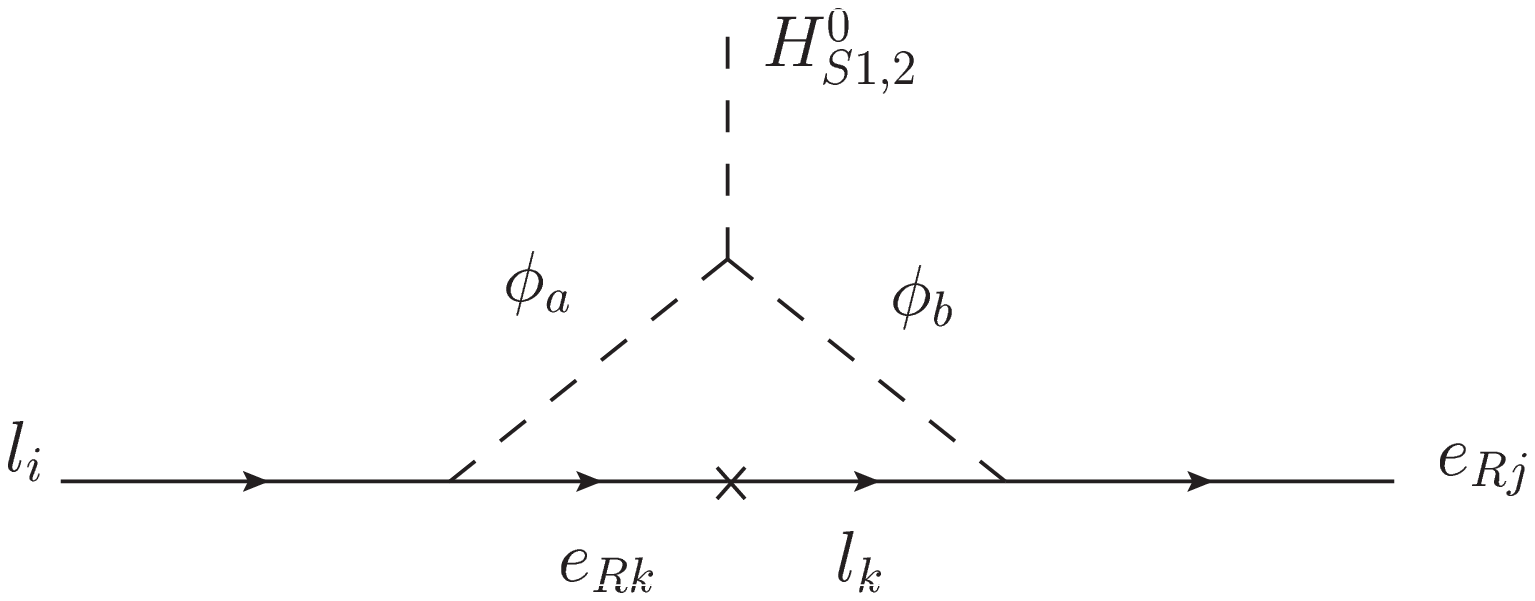,width=\hsize}
   \end{center}
  \end{minipage}
 \end{tabular}
 \caption{1-loop processes which result in change of Yukawa couplings. 
          When ${\cal T}$-trivial scalar bosons $H^0_{S1,2}$ get VEVs, 
          these processes give mass mixing terms drawn in Fig.4. 
 \label{fig;yukawachange}}
\end{figure} 

These Yukawa couplings are vanishing in the ${\cal T}$-conserving limit, 
so that they are suppressed by the ${\cal T}$-breaking terms. 
We can expect that $(e,\, \mu)$ and  $(e,\, \tau)$ elements may be enhanced 
because of the sizable Yukawa couplings, 
as we have seen in the $\mu \to e \gamma$ and $\tau \to e \gamma$ processes. 
Assuming $(\delta m^2_H)_{ee}$ is only nonzero among the ${\cal T}$-breaking terms, 
we find that $y^1_{e \mu }$ is approximately estimated as, 
\beq
y^1_{e \mu} \approx - \frac{ 2 \sqrt 2 C_7}{1-F(m^2_{\phi_e}/m^2_\tau)} \times  \left ( \cos \alpha \frac{\pa m^2_{\phi_e}}{ \pa \langle H_q \rangle} +\sin \alpha \frac{\pa  m^2_{\phi_e}}{ \pa \langle H_1 \rangle} \right ),
\eeq
where the dependence of $\langle H_1 \rangle$ and $\langle H_q \rangle$ in $(\delta m^2_H)_{ee}$ is ignored.
Eventually $y^1_{e \mu}$ is very tiny, 
because of the stringent $\mu \to e \gamma$ constraint. 
When the last term is around the EW-scale, $y^1_{e \mu}$ is at most $O(10^{-12})$. 
$y^1_{e \tau}$ is also small due to the analogy. 
The bound is weaker, so it could be slightly larger than the $(e,\, \mu)$ element, 
but it is at most $O(10^{-9})$. 
The other elements are much smaller, 
because of the suppression of Yukawa couplings and ${\cal T}$-breaking terms. 

\subsection*{contributions to neutrino mixing angles}

Based on the above estimation, 
we investigate the contribution to the LFV, 
and the observed neutrino mixing. 

In many flavor models, the full flavor symmetry is broken to
its subgroups, which are different from each other
between the charged lepton sector and the neutrino sector.
In a certain type of models $\theta_{13} =0$ is predicted at tree level,
while other models lead to non-zero $\theta_{13}$.
Here, we restrict ourselves to the former case, that is,
$\theta_{13} = 0$ at the first stage.
However, as we disscussed above, T-breaking effects from  
$\Delta V\neq0$
may modify the prediction, and neutrino mixing matrix 
is altered to have non-zero $\theta_{13}$.  
Now, we study the contributions of the ${\cal T}$-breaking terms to the neutrino mixing
and discuss the possibility that the observed neutrino mixing is achieved 
by the diagonalizing matrix of charged leptons. 

${\cal S}$-breaking entering into the neutrino sector also gives 
non-zero $\theta_{13}$, but this effect highly depends on 
the setup of neutrino sector; 
whether the right-handed neutrinos is present or not, 
how many if there is, whether the seesaw mechanism arises or not, 
which type of them if it arises, {\it etc.}. 
Thus, we concentrate on the charged lepton sector 
to give model independent considerations. 

The size of correction 
for the off-diagonal elements of Dirac mass matrix is given in Eq. (\ref{eq;neutrino-mixing}).
In addition, $\epsilon_{ij}$ may be induced by extra heavy particles decoupling at
some scale ($\Lambda$) or small deviations of the vacuum alignment, {\it i.e.},
$\langle H^{0}_{e,\mu} \rangle \neq 0$.
Then the diagonalizing matrices $U_L, U_R$ for charged leptons are corrected to be, 
\begin{eqnarray}
U_L^\dag\simeq\left(
\begin{array}{ccc}
1 & -\frac{\epsilon_{e\mu}}{m_\mu} & -\frac{\epsilon_{e\tau}}{m_\tau} \\
\frac{\epsilon_{e\mu}}{m_\mu} & 1 & -\frac{\epsilon_{\mu\tau}}{m_\tau} \\
\frac{\epsilon_{e\tau}}{m_\tau} & \frac{\epsilon_{\mu\tau}}{m_\tau} & 1 
\end{array}\right),~~~~
U_R\simeq\left(
\begin{array}{ccc}
1 & \frac{\epsilon_{\mu e}}{m_\mu} & \frac{\epsilon_{\tau e}}{m_\tau} \\
-\frac{\epsilon_{\mu e}}{m_{\mu}} & 1 & \frac{\epsilon_{\tau\mu}}{m_\tau} \\
-\frac{\epsilon_{\tau e}}{m_\tau} & -\frac{\epsilon_{\tau\mu}}{m_\tau} & 1 
\end{array}\right). \nonumber\\
\label{ULUR}
\end{eqnarray} 
These small deviations modify the neutrino mixing matrix from the Tri-Bi maximal matrix;
\begin{eqnarray}
U_{PMNS}=
\left(
\begin{array}{ccc}
1 & -\frac{\epsilon_{e\mu}}{m_\mu} & -\frac{\epsilon_{e\tau}}{m_\tau} \\
\frac{\epsilon_{e\mu}}{m_\mu} & 1 & -\frac{\epsilon_{\mu\tau}}{m_\tau} \\
\frac{\epsilon_{e\tau}}{m_\tau} & \frac{\epsilon_{\mu\tau}}{m_\tau} & 1 
\end{array}
\right)
\left(
\begin{array}{ccc}
\sqrt{\frac{2}{3}} & \frac{1}{\sqrt{3}} & 0 \\
-\frac{1}{\sqrt{6}} & \frac{1}{\sqrt{3}} & -\frac{1}{\sqrt{2}} \\
-\frac{1}{\sqrt{6}} & \frac{1}{\sqrt{3}} & \frac{1}{\sqrt{2}}
\end{array}
\right). 
\label{eq:PMNS}
\end{eqnarray}
Then $\sin{\theta_{13}}\sim \epsilon_{e\mu}/\sqrt{2}m_{\mu}$ is generated. 
If the contribution of the neutrino sector to $\sin \theta_{13}$ is negligible,
we find the required value for the observed $\sin{\theta_{13}}$:
$\epsilon_{e\mu} \approx 0.2 \times m_\mu$. 

First, let us discuss the one-loop corrections to the neutrino mixings. 
We can write $\theta_{13}$ in terms of ${\cal T}$-breaking effects, according to Eq. (\ref{ULUR}):
\begin{eqnarray}
\sin{\theta_{13}^{\rm lepton}}&=&\frac{\epsilon_{e\mu}}{\sqrt{2}m_\mu}-\frac{\epsilon_{e\tau}}{\sqrt{2}m_\tau}\nonumber\\
                              &\simeq&\frac{Y_{e}^{e\tau}Y_{e}^{\tau\mu}}{\sqrt{2}(4\pi)^2}\frac{m_\tau}{m_\mu}\left(U_{ea}^hU_{ea}^h\ln{\frac{m_{h_a}^2}{m_\tau^2}}-U_{ea}^AU_{ea}^A\ln{\frac{m_{A_a}^2}{m_\tau^2}}\right). 
\end{eqnarray}
Here, we neglect a term of $\epsilon_{e\tau}/m_\tau$, because 
this term is suppressed by a factor $m_\mu^2/m_\tau^2$ 
as compared with the first term. 

\begin{figure}[h]
 \begin{center}
  \includegraphics[width=75mm,angle=270]{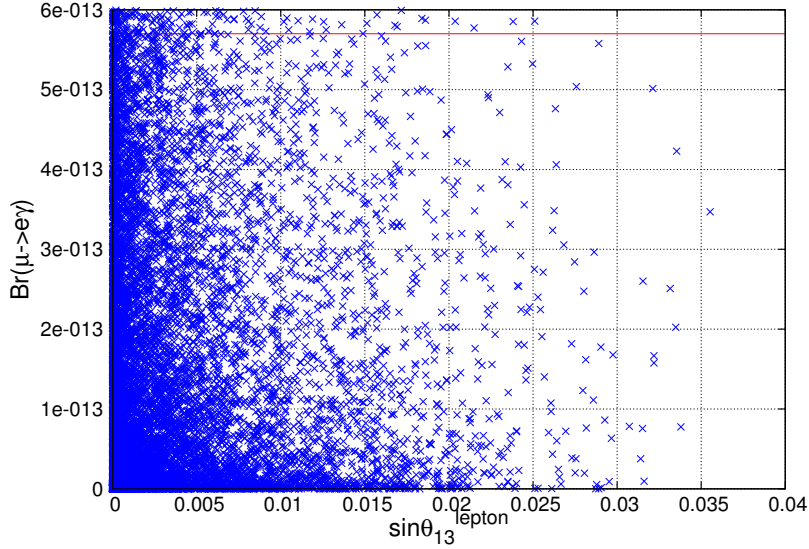}
 \end{center}
  \caption{ 
  Some of possible values of 
           ($\sin{\theta_{13}^{\rm lepton}}$, Br($\mu\to e\gamma$))
           for $\tan{\beta}<100$. 
           The larger $\sin{\theta_{13}^{\rm lepton}}$ is, 
           the smaller value of Br($\mu\to e\gamma$) is tend to favored and vice versa. 
           This is because Br($\mu\to e\gamma$) is suppressed by factors 
           $\frac{1}{m_h^2}$, $\frac{1}{m_A^2}$ 
           compared with $\sin{\theta_{13}^{\rm lepton}}$. 
          The horizontal line shows experimental upper bound 
           of Br($\mu\to e\gamma$) \cite{Adam:2013mnn}.}
 \label{fig:theta13vsbr(mutoegamma)}
\end{figure}

$\epsilon_{ij}$ are generated by the ${\cal T}$-breaking terms which are strongly constrained by especially 
$\mu \to e \gamma$. 
We see the predicted points on $\theta_{13}^{\rm lepton}$ allowed by the $\mu \to e \gamma$ constraint, 
in Fig. \ref{fig:theta13vsbr(mutoegamma)}. 
Note that $\sin{\theta_{13}^{\rm lepton}}$ is proportional to $\cos^{-2}{\beta}$, 
so that it is easily enhanced by $\tan{\beta}$. 
In Fig. \ref{fig:theta13vsbr(mutoegamma)}, $\tan \beta$ is less than $100$, and then $\sin \theta_{13}$
could be $O(0.01)$.
%------------------------------------------------------------------------------
%------------------------------------------------------------------------------
Similary, other shifts of mixing angles from Tri-Bi maximal values 
are written as
\begin{eqnarray}
&&\sin{\theta_{12}^{\rm lepton}}-\frac{1}{\sqrt{3}}\simeq\frac{Y_{e}^{e\tau}Y_{e}^{\tau\mu}}{\sqrt{3}(4\pi)^2}\frac{m_\tau}{m_\mu}\left(U_{ea}^hU_{ea}^h\ln{\frac{m_{h_a}^2}{m_\tau^2}}-U_{ea}^AU_{ea}^A\ln{\frac{m_{A_a}^2}{m_\tau^2}}\right), \nonumber\\
&&\sin{\theta_{23}^{\rm lepton}}-\left(-\frac{1}{\sqrt{2}}\right)\simeq\frac{Y_{e}^{\mu e}Y_{e}^{e\tau}}{\sqrt{2}(4\pi)^2}\frac{m_e}{m_\tau}\left(U_{ea}^hU_{ea}^h\ln{\frac{m_{h_a}^2}{m_e^2}}-U_{e a}^AU_{e a}^A\ln{\frac{m_{A_a}^2}{m_e^2}}\right). \nonumber\\
\end{eqnarray}
$\epsilon_{e\tau}$ in $\sin\theta_{12}$ is neglected 
as in the case of $\theta_{13}$. 
A contribution to $\sin{\theta_{23}^{\rm lepton}}$ 
from ${\cal T}$-breaking in the charged lepton sector 
is extremely small 
because of the factor $m_e/m_\tau$. 
$\sin{\theta_{12}^{\rm lepton}}$ correlates with Br($\mu\to e\gamma$) 
as same as the case of $\theta_{13}^{\rm lepton}$. 
On the other hand, the correction to $\sin{\theta_{23}^{\rm lepton}}$ seems to be much smaller than the others. 

In the above argument, it is attempted to give $\sin\theta_{13}$ 
without considering details of neutrino sector which is highly model dependent. 
However, a contribution to $\sin{\theta_{13}}$ 
coming from only loop-induced ${\cal T}$-breaking effect 
on charged lepton mixing may be small compared to the observed value, 
and then we would like to refer to a possibility of 
other contributions to $\theta_{13}$ from charged lepton sector. 

We can introduce new higher dimensional operators which predict new lepton mixing 
to supply additional contribution to $\theta_{13}$. 
Such operators occur when heavy particles coupling with ${\cal G}$-charged scalars decouple at some scale.
It would be enough to consider additional terms as follows:
\beq
\label{zetaterm}
\zeta_{ijk}\frac{\Phi_i}{\Lambda} H_q \overline{l^j_L}l^k_R+{\rm h.c.}, 
\eeq
where $\Lambda$ is the decoupling scale and $\zeta_{ijk}$ is defined by ${\cal G}$. 
Then, mass mixing terms, $\epsilon_{ij}$, are enhanced, according to the nonzero VEVs of $\Phi$ and $H_q$. 
Especially, the term in Eq. (\ref{zetaterm}) corresponding to $\epsilon_{e\mu}$ adds a new contribution in the form of 
$\zeta\langle\Phi\rangle\langle H_q \rangle/\sqrt{2}m_\mu\Lambda$ 
to $\sin\theta_{13}^{\rm lepton}$. 
Then the size of the coefficient of this effective interaction has to be 
$\zeta/\Lambda=O(10^{-5})\times\langle\Phi\rangle^{-1}$ 
to realize $\sin\theta_{13}=O(0.1)$. 

Secondly, we could consider the possibility that the Higgs VEV alignment is deviated. 
When the VEV alignment of $H_i$ is altered, the remnant symmetry ${\cal T}$ is broken 
and becomes just approximate symmetry even in the charged lepton sector. 
When $H_e^0$ and $H_\mu^0$ gain nonzero VEVs as $\langle H_{e, \, \mu}^0 \rangle= \delta v_{e,\mu}/ \sqrt{2}$, 
mass mixing terms in the form of 
\begin{eqnarray}
(\epsilon^{\delta v}_e)_{e \tau}\equiv Y^{e\tau}_{e}\frac{\delta v_e}{\sqrt{2}}=\frac{m_\tau \delta v_e}{v\cos\beta} ,~ (\epsilon^{\delta v}_\mu)_{e\mu}\equiv Y^{e\mu}_{\mu}\frac{\delta v_\mu}{\sqrt{2}}=\frac{m_\mu\delta v_\mu}{v\cos\beta} 
\label{eq:epsilondeltav}
\end{eqnarray}
are added to $\epsilon_{e\tau}$ and $\epsilon_{e\mu}$ respectively. 
The size of $\delta v_{e, \mu}$ has to be 
$O(0.1)\times v\cos\beta$ to realize $\sin\theta_{13}=O(0.1)$. 
These deviations change the Yukawa couplings for $\phi_{e}$ and $\phi_{\mu}$ 
in the base that charged leptons are mass eigenstates. 
Besides, the mass mixing terms between 
${\cal T}$ -trivial and ${\cal T}$-charged scalars 
can be induced by, for instance, $|H_i|^2|H_q|^2$ term in the Higgs potential. 
Let us approximately describe the deviated Yukawa couplings for $\phi_{e}$ or $\phi_{\mu}$ assuming $\delta v_{e, \mu} \ll v \cos \beta$:
\begin{eqnarray}
\label{eq;additionalYukawa}
{\cal L}'_{\cal T} &=&-Y'^{ij}_e \phi'_e \overline{l^i_L} l^j_R-Y'^{ij}_\mu \phi'_\mu \overline{l^i_L} l^j_R,  \\
Y'^{ij}_e&=& (U^{\dagger}_L)^{ik}Y^{kl}_{e}(U_R)^{lj} - \frac{\delta v _e}{v \cos \beta} \frac{\sqrt{2}m_i}{v \cos \beta } \delta_{ij}, \\
Y'^{ij}_\mu&=& (U^{\dagger}_L)^{ik}Y^{kl}_{\mu}(U_R)^{lj} - \frac{\delta v _\mu}{v \cos \beta} \frac{\sqrt{2}m_i}{v \cos \beta }  \delta_{ij}.
\end{eqnarray}
In these descriptions, we take the SM limit that the SM Higgs around $125$ GeV
does not have tree-level FCNCs and its Yukawa couplings are the same as the SM ones.
Then, the mass bases of $\phi_{e}$ and $\phi_{\mu}$ are slightly deviated by 
$\delta v_e$ and $\delta v_\mu$. The mass bases of CP-even and CP-odd scalars in $\phi'_{e}$ and $\phi'_{\mu}$
may be the same in this limit.

As we have already discussed, the LFV given by $\phi_\mu$ exchanging is dominated by the LFV $\tau$ decay, $\tau^- \to e^+ \mu^- \mu^-$, which is the ${\cal T}$-conserving process. 
The extra ${\cal T}$-breaking terms in addition to the mass mixing in Eq. (\ref{deltam}) 
enhance the other LFV $\tau$ decays as follows: 
\begin{eqnarray}
{\rm Br} (\tau^- \to e^- \mu^+ \mu^- )&\simeq&\left( \frac{\delta v_e}{ v \cos \beta}\frac{m^2_{\phi_\mu}}{m^2_{\phi_e}} \right )^2  \times {\rm Br} (\tau^- \to e^+ \mu^- \mu^-), \\
{\rm Br} (\tau^- \to e^- e^+ \mu^- )&\simeq&\left \{ \left(\frac{\epsilon_{e\tau}}{m_\tau}+\frac{\epsilon_{\tau\mu}}{m_\mu} \right)\frac{m^2_{\phi_\mu}}{m^2_{\phi_e}}+\frac{\epsilon_{e\mu}}{m_\mu} \right \}^2  \times {\rm Br} (\tau^- \to e^+ \mu^- \mu^-),  \\
{\rm Br} (\tau^- \to \mu^- \mu^+ \mu^- )&\simeq& \left( \frac{\epsilon_{e \mu}-\epsilon_{\tau\mu}}{ m_\mu}  - \frac{\delta v_\mu}{ v \cos \beta} \right )^2  \times {\rm Br} (\tau^- \to e^+ \mu^- \mu^-). 
\end{eqnarray}
Note that $\epsilon_{ij}$ in these equations include the contributions of 
the loop corrections, the higher-dimensional operators in Eq. (\ref{zetaterm}), 
and $(\epsilon^{\delta v}_{e, \mu})_{ij}$ in Eq. (\ref{eq:epsilondeltav}). 
The suppression factors in these processes could be estimated as $\sin^2 \theta^{\rm lepton}_{13}$,
so that they are predicted around the region with $ O(10^{-2}) \times {\rm Br} (\tau^- \to e^+ \mu^- \mu^-)$,
which is safe for the current experimental bound
as far as the $\tau^- \to e^+ \mu^- \mu^-$ bound is evaded.

If the CP-even and CP-odd scalars in especially $ \phi'_e$ have different masses, 
the branching ratio of $\mu \to e \gamma$ would be as discussed in Sec. \ref{sec:T-breaking}.
Moreover, the ones of  $\tau \to e \gamma$ and  $\tau \to \mu \gamma$ would be also enhanced,
according to the nonzero $\sin \theta^{\rm lepton}_{13}$:
\begin{eqnarray}
{\rm Br} (\tau^- \to e^- \gamma )&\simeq& \left( \frac{\epsilon_{e \tau}}{ m_\tau}  - \frac{\delta v_e}{ v \cos \beta} \right )^2  \times \frac{m^2_{\tau}}{m^2_\mu}\times {\rm Br} (\mu^- \to e^- \gamma),   \\
{\rm Br} (\tau^- \to \mu^- \gamma )&\simeq& \left( \frac{\epsilon_{e \tau}}{ m_\tau}  - \frac{\delta v_e}{ v \cos \beta} \right )^2  \times  {\rm Br} (\mu^- \to e^- \gamma).
\end{eqnarray}
Eventually, ${\rm Br} (\tau^- \to e^- \gamma )$ is predicted around the region compatible with 
${\rm Br} (\mu^- \to e^- \gamma )$, and then the constraints from the exotic $\tau$ decay
are less serious than the one from $\mu^- \to e^- \gamma$. 

\section{summary}
\label{section6}

The origin of the flavor structure of the fermions in the SM is 
one of the mysteries which have been discussed for a long time.
The SM gauge groups are orthogonal to the generation, and we can find flavor symmetry
rotating the generations, if we ignore the Yukawa couplings to 
generate the mass matrices for the fermions according to the spontaneous
EW symmetry breaking. This fact may suggest the possibility that
the flavor symmetry exists at the high energy and then the observed mass matrices
are generated dynamically. Besides,
we may be able to find some fragments of the flavor symmetry in the SM.
The lepton sector especially may still hold some remnant symmetry of the flavor symmetry
respected at the high-energy scale. The remnant ones
may give some hints for not only model building but also how to prove 
flavor symmetric models in flavor physics.

In this paper, we investigated flavor physics in models with flavor symmetry, ${\cal G}$, 
in a quite general manner. In our setup, only leptons are charged under ${\cal G}$ and
extra Higgs doublets are introduced to respect the flavor symmetry in the Yukawa couplings. 
The Higgs doublets are assumed to belong to non-trivial irreducible representations of ${\cal G}$. 
Then ${\cal G}$ is spontaneously broken 
by VEVs of ${\cal G}$-triplet Higgs bosons, $H_i$ and $\Phi$. 
Some remnant symmetry is left after the symmetry breaking:
${\cal T}$ is conserved in the charged lepton sector and ${\cal S}$ in the neutrino sector respectively. 
This framework has been used to realize a specific neutrino mixing pattern 
such as the Bi maximal or the Tri-Bi maximal mixing. 
The leptophilic ${\cal G}$-triplet $H_i$ breaks ${\cal G}$ to ${\cal T}$, and 
the EW singlet $\Phi_i$, which couples only neutrinos, breaks ${\cal G}$ to ${\cal S}$.

The symmetry ${\cal T}$ plays a crucial role in the control of the FCNCs,
although it is not respected in the full lagrangian.
${\cal T}$-breaking terms would appear in the Higgs potential and neutrino mass matrix,
but they could be also under control once we assume the vacuum alignment of $H_i$ and  $\Phi_i$. 

In our study, ${\cal T}$ is considered as especially ${\cal T}=Z_3$,
and the constraints from the LFV processes are investigated.
In the basis in which ${\cal T}$-generator is diagonal, 
charged leptons are mass eigenstates, 
while details of Higgs potential need to be analyzed 
to decide mass eigenstates of Higgs bosons. 
Charged leptons and mass eigenstates of leptophilic Higgs bosons 
can be classified into trivial and non-trivial ${\cal T}$ singlets after the ${\cal G}$ symmetry breaking,
where the trivial ${\cal T}$-singlet Higgs fields only develop the VEVs.
Then ${\cal T}$ charge conservation constrains 
the form of interactions involving charged leptons: 
the Yukawa couplings of charged leptons and scalars are decided 
by the remnant symmetry. 
Especially, ${\cal T}$-charged Higgs bosons can have FCNCs and 
cause multi-leptonic decays and flavor non-universal gauge couplings. 

We considered the scenario that $\tau$ and $\mu$ leptons carry
non-trivial ${\cal T}$ charges, and then the LFV $\tau$ decay, $\tau^- \to e^+ \mu^- \mu^- $,
is predicted through the ${\cal T}$-charged scalar exchanging. 
The flavor violating scattering, $e^+e^- \to \tau^+ \tau^-$, could be also sizable,
so we investigated the constraints. The masses of ${\cal T}$-charged scalars are expected to be 
around the EW scale, so we conclude that $\tan \beta$ should be less than $O(10)$.

On the other hand, 
the neutrinophilic scalar, $\Phi_i$, breaks ${\cal T}$ in the neutrino sector, 
and this ${\cal T}$ breaking would propagate into the charged lepton sector 
through the interactions between $H_i$ and $\Phi_i$ in the Higgs potential. 
In fact, flavor changing processes that do not conserve ${\cal T}$ charges occur at the one-loop level
such as $\mu\to e\gamma$ and $\tau\to e\gamma$, 
which are important when the model prediction is compared with the experimental bounds. 
In addition, the muon anomalous magnetic moment could be enhanced, although 
the large enhancement is excluded by the bound from the exotic $\tau$ decay. 
Figs.  \ref{fig3} and \ref{fig4} show that $\mu \to e \gamma$ and 
the electron EDM strongly constrain our models if ${\cal T}$ breaking terms in the Higgs potential
are allowed: $\tan \beta  \lesssim 10$. 
In other words, the observations, as well as $\tau^- \to e^+ \mu^- \mu^- $, are the most relevant to
our flavor models. 

In this type of models argued in this paper, 
neutrino mixing with $\theta_{13}=0$ tends to be predicted 
because Bi maximal or Tri-Bi maximal mixing is realized. 
Then the remnant symmetry breaking effects may be required to alter this mixing pattern. 
In Sec. \ref{section5}, 
we discussed the case that the remnant symmetry ${\cal T}$ is slightly broken in the charged lepton sector. 
As mentioned above, the ${\cal T}$ breaking effect caused by VEVs of $\Phi_i$ 
enters into the charged lepton sector through the loop processes 
involving scalars. 
However, this contribution is too small to achieve the observed value of $\theta_{13}$. 
We also considered additional ${\cal T}$-breaking effects to realize the large $\theta_{13}$. 
Such newly added ${\cal T}$-breaking terms also contributes to FCNCs,  
then we pointed out the correlation 
between $\theta_{13}$ and FCNCs caused by additional ${\cal T}$ breaking effect. 
For instance, the branching ratios of $\tau^- \to e^- \mu^+ \mu^- $, $e^- e^+ \mu^-$ and $\mu^- \mu^+ \mu^-$ becomes the almost same order as Br$(\tau^- \to e^+ \mu^- \mu^- )$ suppressed by $O(10^{-2})$,
and Br($\tau^- \to e^- \gamma$) may be the same order as Br($\mu^- \to e^- \gamma$).
The process, $\tau^- \to e^+ \mu^- \mu^-$, is the most important in our model,
but these predictions would be also useful to test our flavor models.

In this study, we take the SM limit, so the SM-like Higgs around $125$ GeV does not have FCNCs.
As discussed recently in Refs. \cite{Omura, hmutau}, it may be interesting to allow the tree-level FCNCs involving 
the SM-like Higgs, motivated by the CMS excess in the $h \to \mu \tau$ channel \cite{Khachatryan:2015kon} as well as the 
muon anomalous magnetic moment \cite{Omura}. However, our model would not enhance the $(g-2)_{\mu}$ because
of the chirality structure of the FCNCs,
and then the possible way is to consider very light pseudoscalar and large $\tan \beta$
to achieve the discrepancy of $(g-2)_{\mu}$ \cite{leptophilic2HDM}. Such a parameter set
would require the tuning of the ${\cal T}$-breaking terms and large
mass differences among the scalars to evade the strong bounds from $\mu \to e \gamma$, the electron EDM and $\tau^- \to e^+ \mu^- \mu^-$.

Note that we studied flavor physics assuming the relation in Eq. (\ref{A4-relation})
and ${\cal T}=Z_n$ ($n\neq 2$). 
These conditions may be relevant to our prediction, so we will
discuss the other possibility such as ${\cal T}=Z_2$ in the future. 

%---------------------------------------------------------------------------
\acknowledgments
%---------------------------------------------------------------------------
This work is supported by Grant-in-Aid for Scientific research
from the Ministry of Education, Science, Sports, and Culture (MEXT),
Japan, No. 23104011 (for Y.O.) and No. 25400252 (for T.K.).

\appendix
\section{Mass Matrices and vacuum alignment in $A_4$ model}

As an illustration of the argument in Sec. \ref{section3}, 
I show the most simple example with ${\cal G}=A_4$, ${\cal T}=Z_3$, ${\cal S}=Z_2$. 
In this model, $H_q$ is $A_4$ singlet, 
$H=(H_1, H_2, H_3)$ and $\Phi=(\Phi_1, \Phi_2, \Phi_3)$ are $A_4$ triplets. 
$H_q$ and $H_i~(i=1,2,3)$ are $SU(2)_L$ doublets, 
and $\Phi_i$ are treated as gauge singlet real scalar bosons for simplisity.

In the base that diagonalize $Z_3$ generator $T_H$, $A_4$ is generated by, 
\begin{eqnarray}
T_H=\left(
\begin{array}{ccc}
1 & 0 & 0 \\
0 & \omega & 0 \\
0 & 0 & \omega^2
\end{array}
\right),~~
S_\Phi=\frac{1}{3}\left(
\begin{array}{ccc}
-1 & 2 & 2 \\
2 & -1 & 2 \\
2 & 2 & -1 
\end{array}
\right), 
\end{eqnarray}
for {\bf 3} representation. 
The generators for {\bf 1}, {\bf 1'} and {\bf 1''} representations of $A_4$ are $S_\Phi= {\bf 1}$ and 
$T_H= {\bf1},$ $\omega$, $\omega^2$ respectively. 

In the base in which ${\cal S}$ generator is diagonal,
two generators of $A_4$ are written as
\begin{eqnarray}
\widetilde S_\Phi=\left(
\begin{array}{ccc}
1 &  0 &  0 \\
0 & -1 &  0 \\
0 &  0 & -1
\end{array}
\right),~~~~
\widetilde T_H=\left(
\begin{array}{ccc}
0 & 1 & 0 \\
0 & 0 & 1 \\
1 & 0 & 0 
\end{array}
\right), 
\label{Z3diag}
\end{eqnarray}
for ${\bm 3}$ representation. 
For trivial and non-trivial singlets, these two generators are $\widetilde S_\Phi= {\bf 1}$, $ \widetilde T_H= {\bf 1},~\omega,~\omega^2$. 
$\Phi$ is here assumed to be gauge singlet for simplicity. 
In this basis, the general form of interaction terms between scalar bosons 
in our $A_4$ model, which can give the expected vacuum alignment, 
\footnote{We do not include three point interaction terms 
in the Higgs potential below because they cause tadpole terms 
of ${\cal T}$ and ${\cal S}$-charged fields which make our VEV alignment unstable. }
is, 
\begin{eqnarray}
V_H(H_q, H_i)&=&\mu_q^2H_q^\dag H_q+\mu^2\{H_1^\dag H_1+H_2^\dag H_2+H_3^\dag H_3\} \nonumber \\
             &&+\lambda_1(H_q^\dag H_q)^2+\lambda_2\{(H_1^\dag H_1)^2+(H_2^\dag H_2)^2+(H_3^\dag H_3)^2\} \nonumber \\
             &&+\lambda_3\{(H_1^\dag H_1)(H_2^\dag H_2)+(H_2^\dag H_2)(H_3^\dag H_3)+(H_3^\dag H_3)(H_1^\dag H_1)\} \nonumber \\
             &&+\lambda_4\{(H_1^\dag H_2)^2+(H_2^\dag H_3)^2+(H_3^\dag H_1)^2+{\rm h.c.}\}+\lambda_5\{|H_1^\dag H_2|^2+|H_2^\dag H_3|^2+|H_3^\dag H_1|^2\} \nonumber \\
             &&+\lambda_6\{(H_1^\dag H_1)(H_q^\dag H_q)+(H_2^\dag H_2)(H_q^\dag H_q)+(H_3^\dag H_3)(H_q^\dag H_q)\} \nonumber \\
             &&+\lambda_7\{(H_1^\dag H_q)^2+(H_2^\dag H_q)^2+(H_3^\dag H_q)^2+{\rm h.c.}\}+\lambda_8\{|H_1^\dag H_q|^2+|H_2^\dag H_q|^2+|H_3^\dag H_q|^2\} \nonumber \\
             &&+\lambda_9\{(H_1^\dag H_2)(H_3^\dag H_q)+(H_2^\dag H_3)(H_1^\dag H_q)+(H_3^\dag H_1)(H_2^\dag H_q)+{\rm h.c.}\} \nonumber \\
             &&+\lambda_{10}\{(H_2^\dag H_1)(H_3^\dag H_q)+(H_3^\dag H_2)(H_1^\dag H_q)+(H_1^\dag H_3)(H_2^\dag H_q)+{\rm h.c.}\}, \\
V_\Phi(H_q, \Phi_i)&=&\frac{\mu_\Phi^2}{2}(\Phi_1^2+\Phi_2^2+\Phi_3^2)\nonumber\\
                     &&+\lambda_1^\Phi(\Phi_1^4+\Phi_2^4+\Phi_3^4)+\lambda_2^\Phi(\Phi_1^2\Phi_2^2+\Phi_2^2\Phi_3^2+\Phi_3^2\Phi_1^2) \nonumber\\
                     &&+\lambda_3^\Phi H_q^\dag H_q (\Phi_1^2+\Phi_2^2+\Phi_3^2), \\
\Delta V(H_i, \Phi_i)&=&\lambda_1^\Delta(\Phi_1^2H_1^\dag H_1+\Phi_2^2H_2^\dag H_2+\Phi_3^2H_3^\dag H_3) \nonumber\\
                      &&+\lambda_2^\Delta(\Phi_1^2H_2^\dag H_2+\Phi_2^2H_3^\dag H_3+\Phi_3^2H_1^\dag H_1)+\lambda_3^\Delta(\Phi_1^2H_3^\dag H_3+\Phi_2^2H_1^\dag H_1+\Phi_3^2H_2^\dag H_2) \nonumber\\
                      &&+\lambda_4^\Delta(\Phi_1\Phi_2H_1^\dag H_2+\Phi_2\Phi_3H_2^\dag H_3+\Phi_3\Phi_1H_3^\dag H_1+{\rm h.c.}) 
\label{eq:potential2}
\end{eqnarray}

In order to leave $Z_3$ in the charged lepton sector and $Z_2$ in the neutrino sector, 
leptophilic boson $H$ and neutrinophilic boson $\Phi$ have to obtain VEVs as 
\beq
(\langle H_1 \rangle, \langle H_2 \rangle, \langle H_3 \rangle)=(v_H/\sqrt 2, v_H/\sqrt 2, v_H/\sqrt 2),~~
(\langle \Phi_1 \rangle, \langle \Phi_2 \rangle, \langle \Phi_3 \rangle)=(v_\Phi/\sqrt 2, 0, 0). 
\eeq
$H_q$ also get VEV $v_q$ to give quark masses. 
Thus, we expand Eq. (\ref{eq:potential2}) around the vacuum, 
\begin{eqnarray}
H_q=\left(\begin{array}{c} H_q^+ \\ \frac{1}{\sqrt{2}}(v_q+H_q^0+iA_q) \end{array}\right),~~~H_j=\left(\begin{array}{c} H_j^+ \\ \frac{1}{\sqrt{2}}(v_H+H_j^0+iA_j) \end{array}\right)~~(j=1,2,3)
\end{eqnarray} 
and $\Phi_1=v_\Phi/\sqrt 2+\phi_1, \Phi_{2, 3}=\phi_{2,3}$ 
to derive mass matrix for scalar bosons.
We here assume that the effect of $\Delta V$ is negligibly small 
as noted in Sec. \ref{section3}, that is, $\lambda_j^\Delta \sim 0$. 
The mass matrices for charged scalar bosons ($H_q^+, H_1^+, H_2^+, H_3^+$) 
and for neutral CP-odd scalar bosons $(A_q, A_1, A_2, A_3)$ in the $S_\Phi$-diagonal base are
\begin{eqnarray}
M^2=\left(\begin{array}{cccc}
a & b & b & b \\
b & c & d & d \\
b & d & c & d \\
b & d & d & c 
\end{array}\right),
\end{eqnarray}
where $a$, $b$, $c$ and $d$ are defined as
\begin{eqnarray}
a&=&\mu_q^2+\lambda_1v_q^2+\frac{3}{2}\lambda_6v_H^2+\lambda_3^\Phi v_\Phi^2, \nonumber\\
b&=&\frac{1}{2}(2\lambda_7+\lambda_8)v_qv_H+\frac{1}{2}(\lambda_9+\lambda_{10})v_H^2, \nonumber\\
c&=&\mu^2+(\lambda_2+\lambda_3)v_H^2+\frac{1}{2}\lambda_6v_q^2, \nonumber\\
d&=&\frac{1}{2}(2\lambda_4+\lambda_5)v_H^2+\frac{1}{2}(\lambda_9+\lambda_{10})v_qv_H,
\end{eqnarray}
for charged scalars, and 
\begin{eqnarray}
a&=&\frac{\mu_q^2}{2}+\frac{1}{2}\lambda_1v_q^2+\frac{3}{4}(\lambda_6-2\lambda_7+\lambda_8)v_H^2+\frac{\lambda_3^\Phi}{2}v_\Phi^2, \nonumber\\
b&=&\lambda_7v_qv_H+\frac{1}{4}(\lambda_9+\lambda_{10})v_H^2, \nonumber\\
c&=&\frac{\mu^2}{2}+\frac{1}{2}(\lambda_2+\lambda_3-2\lambda_4+\lambda_5)v_H^2+\frac{1}{4}(\lambda_6-2\lambda_7+\lambda_8)v_q^2, \nonumber\\
d&=&\lambda_4v_H^2+\frac{1}{4}(\lambda_9+\lambda_{10})v_qv_H,
\end{eqnarray}
for neutral CP-odd states. 

The mass matrix for the neutral CP-even scalar bosons 
$(H_q^0, H_1^0, H_2^0, H_3^0, \phi_1, \phi_2, \phi_3)$  in the $S_\Phi$-diagonal base is
\begin{eqnarray}
M^2=\left(\begin{array}{c|ccc|ccc}
A & B & B & B & G & 0 & 0 \\\hline
B & C & D & D &   &   &   \\
B & D & C & D &   & 0 &   \\
B & D & D & C &   &   &   \\\hline
G &   &   &   & E & 0 & 0 \\
0 &   & 0 &   & 0 & F & 0 \\
0 &   &   &   & 0 & 0 & F
\end{array}\right),
\end{eqnarray}
where each element is defined as follows:
\begin{eqnarray}
A&=&\frac{\mu_q^2}{2}+\frac{3}{2}\lambda_1v_q^2+\frac{3}{4}(\lambda_6+2\lambda_7+\lambda_8)v_H^2+\frac{\lambda_3^\Phi}{2}v_\Phi^2, \nonumber\\
B&=&\frac{1}{2}(\lambda_6+2\lambda_7+\lambda_8)v_qv_H+\frac{3}{4}(\lambda_9+\lambda_{10})v_H^2, \nonumber\\
C&=&\frac{\mu^2}{2}+\frac{1}{2}(3\lambda_2+\lambda_3+2\lambda_4+\lambda_5)v_H^2+\frac{1}{4}(\lambda_6+2\lambda_7+\lambda_8)v_q^2, \nonumber\\
D&=&\frac{1}{2}(\lambda_3+2\lambda_4+\lambda_5)v_H^2+\frac{3}{4}(\lambda_9+\lambda_{10})v_qv_H, \nonumber\\
E&=&\frac{\mu_\Phi^2}{2}+6\lambda_1^\Phi v_\Phi^2+\frac{1}{2}\lambda_3^\Phi v_q^2, \nonumber\\
F&=&\frac{\mu_\Phi^2}{2}+\lambda_2^\Phi v_\Phi^2+\frac{1}{2}\lambda_3^\Phi v_q^2, \nonumber\\
G&=&\lambda_3^\Phi v_\Phi v_q. 
\end{eqnarray}

In order to move to the basis in which $T_H$ is diagonal
we rotate $A_4$ triplets with
\begin{eqnarray}
U_\omega=\frac{1}{\sqrt{3}}\left(
\begin{array}{ccc}
1 & 1 & 1 \\
1 & \omega^2 & \omega \\
1 & \omega & \omega^2
\end{array}
\right), 
\end{eqnarray}
for $U_\omega T_H U_\omega^\dag={\rm diag}(1, \omega, \omega^2)$, 
which is the form of Eq. (\ref{T}), and the VEV is defined as 
\begin{eqnarray}
\sqrt{3}v_H=v\cos{\beta},~~ v_q=v\sin{\beta}. 
\end{eqnarray}
All mass submatrices for $A_4$-triplet scalar bosons are rotated
and then we find mass matrices of the scalars in the form of Eq. (\ref{eq:V''}). 

Nonzero $\Delta V$ causes ${\cal T}$-breaking terms as follows:
\begin{eqnarray}
\delta m^2&=&\frac{1}{2}\lambda_1^\Delta v_\Phi^2 H_1^{02}+\frac{1}{2}\lambda_2^\Delta v_\Phi^2 H_2^{02}+\frac{1}{2}\lambda_3^\Delta v_\Phi^2 H_3^{02} \nonumber\\
           &&+\frac{1}{2}\lambda_1^\Delta v_\Phi^2A_1^2+\frac{1}{2}\lambda_2^\Delta v_\Phi^2A_2^2+\frac{1}{2}\lambda_3^\Delta v_\Phi^2A_3^2 \nonumber\\
           &&+\lambda_1^\Delta v_\Phi^2 H_1^- H_1^+ +\lambda_2^\Delta v_\Phi^2 H_2^- H_2^+ +\lambda_3^\Delta v_\Phi^2 H_3^- H_3^+ \nonumber\\
           &&+\frac{1}{2}(\lambda_1^\Delta+\lambda_2^\Delta+\lambda_3^\Delta)v_H^2(\phi_1^2+\phi_2^2+\phi_3^2)+\lambda_4^\Delta v_H^2 (\phi_1\phi_2+\phi_2\phi_3+\phi_3\phi_1) \nonumber\\
           &&+2\lambda_1^\Delta v_\Phi v_H H_1^0\phi_1+2\lambda_2 v_\Phi v_H H_2^0\phi_1+2\lambda_3^\Delta v_\Phi v_H H_3^0\phi_1  \nonumber \\
           &&+\lambda_4^\Delta v_\Phi v_H(H_1^0\phi_2+H_2^0\phi_2+H_1^0\phi_3+H_3^0\phi_3). 
\label{breakingterms}
\end{eqnarray}

In addition, we can read out the ${\cal T}$ breaking terms in Eq. (\ref{deltam})
for this model from Eq. (\ref{breakingterms}):
\beq
(\delta m^2_H)_{ab}=(\delta m^2_A)_{ab}=(\delta m^2_{H^+})_{ab}=\lambda_a^\Delta v_\Phi^2 \delta_{ab}, 
\eeq
in the basis of Eq. (\ref{Z3diag}). 
After rotated by $U_\omega$, 
mixing terms between states with different $Z_3$ charges occur 
from these $(\delta m^2)_{ab}$. 
These terms contribute to ${\cal T}$-breaking 
in the charged lepton sector via loop corrections. 
Further ${\cal T}$-breaking terms may be required  
to describe observed size of non-zero $\theta_{13}$ 
as we discuss in Sec. \ref{section5}.

%------------------------------------------------------------------------------


\vspace{-1ex}

\begin{thebibliography}{99}

\bibitem{Glashow:1970gm} 
  S.~L.~Glashow, J.~Iliopoulos and L.~Maiani,
  Phys.\ Rev.\ D {\bf 2}, 1285 (1970).

\bibitem{Altarelli} 
  G.~Altarelli and F.~Feruglio,
  Rev.\ Mod.\ Phys.\  {\bf 82}, 2701 (2010)
  [arXiv:1002.0211 [hep-ph]].
  
\bibitem{Ishimori} 
  H.~Ishimori, T.~Kobayashi, H.~Ohki, Y.~Shimizu, H.~Okada and M.~Tanimoto,
  Prog.\ Theor.\ Phys.\ Suppl.\  {\bf 183}, 1 (2010)
  [arXiv:1003.3552 [hep-th]].

\bibitem{S4}
  S.~F.~King and C.~Luhn,
  Rept.\ Prog.\ Phys.\  {\bf 76}, 056201 (2013)
  [arXiv:1301.1340 [hep-ph]].

\bibitem{TB} 
  P.~F.~Harrison, D.~H.~Perkins and W.~G.~Scott,
  Phys.\ Lett.\ B {\bf 530}, 167 (2002)
  [hep-ph/0202074];
  P.~F.~Harrison and W.~G.~Scott,
  Phys.\ Lett.\ B {\bf 535}, 163 (2002)
  [hep-ph/0203209];
  Phys.\ Lett.\ B {\bf 557}, 76 (2003)
  [hep-ph/0302025].

\bibitem{A4}
  E.~Ma and G.~Rajasekaran,
  Phys.\ Rev.\ D {\bf 64}, 113012 (2001)
  [hep-ph/0106291];
  G.~Altarelli and F.~Feruglio,
  Nucl.\ Phys.\ B {\bf 741}, 215 (2006)
  [hep-ph/0512103].

\bibitem{A4-3} 
  G.~Altarelli and D.~Meloni,
  J.\ Phys.\ G {\bf 36}, 085005 (2009)
  [arXiv:0905.0620 [hep-ph]].

\bibitem{flavor2} 
  M.~Hirsch, S.~Morisi, E.~Peinado and J.~W.~F.~Valle,
  Phys.\ Rev.\ D {\bf 82}, 116003 (2010)
  [arXiv:1007.0871 [hep-ph]];
  M.~S.~Boucenna, M.~Hirsch, S.~Morisi, E.~Peinado, M.~Taoso and J.~W.~F.~Valle,
  JHEP {\bf 1105}, 037 (2011)
  [arXiv:1101.2874 [hep-ph]].

\bibitem{S4-0}
  C.~S.~Lam,
  Phys.\ Lett.\ B {\bf 656}, 193 (2007)
  [arXiv:0708.3665 [hep-ph]];
  Phys.\ Rev.\ Lett.\  {\bf 101}, 121602 (2008)
  [arXiv:0804.2622 [hep-ph]];
  Phys.\ Rev.\ D {\bf 78}, 073015 (2008)
  [arXiv:0809.1185 [hep-ph]].

\bibitem{delta27-1} 
  I.~de Medeiros Varzielas, S.~F.~King and G.~G.~Ross,
  Phys.\ Lett.\ B {\bf 648}, 201 (2007)
  [hep-ph/0607045].

\bibitem{theta13-1} 
  F.~P.~An {\it et al.}  [DAYA-BAY Collaboration],
  Phys.\ Rev.\ Lett.\  {\bf 108}, 171803 (2012)
  [arXiv:1203.1669 [hep-ex]];
  Chin.\ Phys.\ C {\bf 37}, 011001 (2013)
  [arXiv:1210.6327 [hep-ex]].

\bibitem{theta13-2} 
  Y.~Abe {\it et al.}  [DOUBLE-CHOOZ Collaboration],
  Phys.\ Rev.\ Lett.\  {\bf 108}, 131801 (2012)
  [arXiv:1112.6353 [hep-ex]].

\bibitem{theta13-3} 
  J.~K.~Ahn {\it et al.}  [RENO Collaboration],
  Phys.\ Rev.\ Lett.\  {\bf 108}, 191802 (2012)
  [arXiv:1204.0626 [hep-ex]].

\bibitem{theta13-4} 
  K.~Abe {\it et al.}  [T2K Collaboration],
  Phys.\ Rev.\ Lett.\  {\bf 107}, 041801 (2011)
  [arXiv:1106.2822 [hep-ex]].

\bibitem{theta13-5} 
  P.~Adamson {\it et al.}  [MINOS Collaboration],
  Phys.\ Rev.\ Lett.\  {\bf 107}, 181802 (2011)
  [arXiv:1108.0015 [hep-ex]].

\bibitem{Hamada:2014xha} 
  Y.~Hamada, T.~Kobayashi, A.~Ogasahara, Y.~Omura, F.~Takayama and D.~Yasuhara,
  JHEP {\bf 1410}, 183 (2014)
  [arXiv:1405.3592 [hep-ph]].

\bibitem{A4-2} 
  G.~Altarelli, F.~Feruglio, L.~Merlo and E.~Stamou,
  JHEP {\bf 1208}, 021 (2012)
  [arXiv:1205.4670 [hep-ph]].

\bibitem{A4-4} 
  Y.~Lin,
  Nucl.\ Phys.\ B {\bf 824}, 95 (2010)
  [arXiv:0905.3534 [hep-ph]].

\bibitem{S4-3} 
  F.~Feruglio, C.~Hagedorn and R.~Ziegler,
  JHEP {\bf 1307}, 027 (2013)
  [arXiv:1211.5560 [hep-ph]].

\bibitem{delta27}
  E.~Ma,
  Phys.\ Lett.\ B {\bf 723}, 161 (2013)
  [arXiv:1304.1603 [hep-ph]].

\bibitem{flavor-higher} 
  H.~Ishimori, Y.~Shimizu, M.~Tanimoto and A.~Watanabe,
  Phys.\ Rev.\ D {\bf 83}, 033004 (2011)
  [arXiv:1010.3805 [hep-ph]];
  S.~F.~King and C.~Luhn,
  JHEP {\bf 1109}, 042 (2011)
  [arXiv:1107.5332 [hep-ph]];
  R.~Krishnan, P.~F.~Harrison and W.~G.~Scott,
  JHEP {\bf 1304}, 087 (2013)
  [arXiv:1211.2000 [hep-ph]];
  Y.~Grossman and W.~H.~Ng,
  arXiv:1404.1413 [hep-ph];
  F.~Feruglio, C.~Hagedorn and R.~Ziegler,
  Eur.\ Phys.\ J.\ C {\bf 74}, 2753 (2014)
  [arXiv:1303.7178 [hep-ph]].

\bibitem{theta13-model}
  T.~Araki and Y.~F.~Li,
  Phys.\ Rev.\ D {\bf 85}, 065016 (2012)
  [arXiv:1112.5819 [hep-ph]];
  S.~Dev, R.~R.~Gautam and L.~Singh,
  Phys.\ Lett.\ B {\bf 708}, 284 (2012)
  [arXiv:1201.3755 [hep-ph]];
  P.~S.~Bhupal Dev, B.~Dutta, R.~N.~Mohapatra and M.~Severson,
  Phys.\ Rev.\ D {\bf 86}, 035002 (2012)
  [arXiv:1202.4012 [hep-ph]];
  I.~K.~Cooper, S.~F.~King and C.~Luhn,
  JHEP {\bf 1206}, 130 (2012)
  [arXiv:1203.1324 [hep-ph]];
  Y.~H.~Ahn and S.~K.~Kang,
  Phys.\ Rev.\ D {\bf 86}, 093003 (2012)
  [arXiv:1203.4185 [hep-ph]];
  I.~de Medeiros Varzielas and G.~G.~Ross,
  JHEP {\bf 1212}, 041 (2012)
  [arXiv:1203.6636 [hep-ph]];
  G.~Altarelli, F.~Feruglio and L.~Merlo,
  Fortsch.\ Phys.\  {\bf 61}, 507 (2013)
  [arXiv:1205.5133 [hep-ph]];
  M.~C.~Chen, J.~Huang, J.~M.~O'Bryan, A.~M.~Wijangco and F.~Yu,
  JHEP {\bf 1302}, 021 (2013)
  [arXiv:1210.6982 [hep-ph]];
  C.~Luhn,
  Nucl.\ Phys.\ B {\bf 875}, 80 (2013)
  [arXiv:1306.2358 [hep-ph]];
  G.~J.~Ding and S.~F.~King,
  Phys.\ Rev.\ D {\bf 89}, no. 9, 093020 (2014)
  [arXiv:1403.5846 [hep-ph]];
  G.~J.~Ding and Y.~L.~Zhou,
  JHEP {\bf 1406}, 023 (2014)
  [arXiv:1404.0592 [hep-ph]].

\bibitem{King} 
  S.~F.~King and C.~Luhn,
  JHEP {\bf 0910}, 093 (2009)
  [arXiv:0908.1897 [hep-ph]].

\bibitem{T7}
  Q.~H.~Cao, S.~Khalil, E.~Ma and H.~Okada,
  Phys.\ Rev.\ Lett.\  {\bf 106}, 131801 (2011)
  [arXiv:1009.5415 [hep-ph]];
  C.~Luhn, K.~M.~Parattu and A.~Wingerter,
  JHEP {\bf 1212}, 096 (2012)
  [arXiv:1210.1197 [hep-ph]].

\bibitem{flavor}
  E.~Ma,
  Phys.\ Rev.\ D {\bf 70}, 031901 (2004)
  [hep-ph/0404199];
  A.~E.~Carcamo Hernandez, I.~de Medeiros Varzielas, S.~G.~Kovalenko, H.~P\"as and I.~Schmidt,
  Phys.\ Rev.\ D {\bf 88}, no. 7, 076014 (2013)
  [arXiv:1307.6499 [hep-ph]];
  M.~Holthausen, M.~Lindner and M.~A.~Schmidt,
  Phys.\ Rev.\ D {\bf 87}, no. 3, 033006 (2013)
  [arXiv:1211.5143 [hep-ph]].

\bibitem{S4-2} 
  G.~Altarelli, F.~Feruglio and L.~Merlo,
  JHEP {\bf 0905}, 020 (2009)
  [arXiv:0903.1940 [hep-ph]].

\bibitem{A5} 
  A.~Di Iura, C.~Hagedorn and D.~Meloni,
  arXiv:1503.04140 [hep-ph];
  P.~Ballett, S.~Pascoli and J.~Turner,
  arXiv:1503.07543 [hep-ph].

\bibitem{delta6n2} 
  R.~d.~A.~Toorop, F.~Feruglio and C.~Hagedorn,
  Phys.\ Lett.\ B {\bf 703}, 447 (2011)
  [arXiv:1107.3486 [hep-ph]];
  R.~de Adelhart Toorop, F.~Feruglio and C.~Hagedorn,
  Nucl.\ Phys.\ B {\bf 858}, 437 (2012)
  [arXiv:1112.1340 [hep-ph]];
  G.~J.~Ding,
  Nucl.\ Phys.\ B {\bf 862}, 1 (2012)
  [arXiv:1201.3279 [hep-ph]];
  S.~F.~King, C.~Luhn and A.~J.~Stuart,
  Nucl.\ Phys.\ B {\bf 867}, 203 (2013)
  [arXiv:1207.5741 [hep-ph]];
  C.~S.~Lam,
  Phys.\ Rev.\ D {\bf 87}, no. 1, 013001 (2013)
  [arXiv:1208.5527 [hep-ph]];
  M.~Holthausen, K.~S.~Lim and M.~Lindner,
  Phys.\ Lett.\ B {\bf 721}, 61 (2013)
  [arXiv:1212.2411 [hep-ph]];
  C.~S.~Lam,
  Phys.\ Rev.\ D {\bf 87}, no. 5, 053012 (2013)
  [arXiv:1301.1736 [hep-ph]];
  S.~F.~King, T.~Neder and A.~J.~Stuart,
  Phys.\ Lett.\ B {\bf 726}, 312 (2013)
  [arXiv:1305.3200 [hep-ph]].

\bibitem{Aoki:2009ha} 
  M.~Aoki, S.~Kanemura, K.~Tsumura and K.~Yagyu,
  Phys.\ Rev.\ D {\bf 80}, 015017 (2009)
  [arXiv:0902.4665 [hep-ph]]. 

\bibitem{Branco:2011iw} 
  G.~C.~Branco, P.~M.~Ferreira, L.~Lavoura, M.~N.~Rebelo, M.~Sher and J.~P.~Silva,
  Phys.\ Rept.\  {\bf 516}, 1 (2012)
  [arXiv:1106.0034 [hep-ph]].

\bibitem{Kobayashi:2008ih} 
  T.~Kobayashi, Y.~Omura and K.~Yoshioka,
  Phys.\ Rev.\ D {\bf 78}, 115006 (2008)
  [arXiv:0809.3064 [hep-ph]].

\bibitem{Abbiendi:2013hk} 
  G.~Abbiendi {\it et al.}  [ALEPH and DELPHI and L3 and OPAL and LEP Collaborations],
  Eur.\ Phys.\ J.\ C {\bf 73}, 2463 (2013)
  [arXiv:1301.6065 [hep-ex]].

\bibitem{LEPbound} 
  t.~S.~Electroweak [LEP and ALEPH and DELPHI and L3 and OPAL and LEP Electroweak Working Group and SLD Electroweak Group and SLD Heavy Flavor Group Collaborations],
  hep-ex/0312023.

\bibitem{Hayasaka:2010np} 
  K.~Hayasaka, K.~Inami, Y.~Miyazaki, K.~Arinstein, V.~Aulchenko, T.~Aushev, A.~M.~Bakich and A.~Bay {\it et al.},
  Phys.\ Lett.\ B {\bf 687}, 139 (2010)
  [arXiv:1001.3221 [hep-ex]].

\bibitem{Bellgardt:1987du}
  U.~Bellgardt {\it et al.}  [SINDRUM Collaboration],
  Nucl.\ Phys.\ B {\bf 299} (1988) 1.

\bibitem{LFV-2HDM} 
  A.~Crivellin, A.~Kokulu and C.~Greub,
  Phys.\ Rev.\ D {\bf 87}, no. 9, 094031 (2013)
  [arXiv:1303.5877 [hep-ph]].

\bibitem{PDG}
K.A. Olive et al. (Particle Data Group), Chin. Phys. C, 38, 090001 (2014). 

\bibitem{Pich:2013lsa} 
  A.~Pich,
  Prog.\ Part.\ Nucl.\ Phys.\  {\bf 75}, 41 (2014)
  [arXiv:1310.7922 [hep-ph]].

\bibitem{leptophilic2HDM} 
  A.~Broggio, E.~J.~Chun, M.~Passera, K.~M.~Patel and S.~K.~Vempati,
  JHEP {\bf 1411}, 058 (2014)
  [arXiv:1409.3199 [hep-ph]];
  T.~Abe, R.~Sato and K.~Yagyu,
  arXiv:1504.07059 [hep-ph].

\bibitem{chargedLHC}
  G.~Aad {\it et al.}  [ATLAS Collaboration],
  arXiv:1412.6663 [hep-ex];
  CMS Collaboration, CMS-PAS-HIG-14-020, CERN, Geneva Switzerland (2014).

\bibitem{Hermann:2012fc} 
  T.~Hermann, M.~Misiak and M.~Steinhauser,
  JHEP {\bf 1211}, 036 (2012)
  [arXiv:1208.2788 [hep-ph]].

\bibitem{LFV-2HDM-2} 
  S.~Davidson and G.~J.~Grenier,
  Phys.\ Rev.\ D {\bf 81}, 095016 (2010)
  [arXiv:1001.0434 [hep-ph]].

\bibitem{Chang:1993kw} 
  D.~Chang, W.~S.~Hou and W.~Y.~Keung,
  Phys.\ Rev.\ D {\bf 48}, 217 (1993)
  [hep-ph/9302267].

\bibitem{Adam:2013mnn} 
  J.~Adam {\it et al.}  [MEG Collaboration],
  Phys.\ Rev.\ Lett.\  {\bf 110}, 201801 (2013)
  [arXiv:1303.0754 [hep-ex]].

\bibitem{Baldini:2013ke} 
  A.~M.~Baldini, F.~Cei, C.~Cerri, S.~Dussoni, L.~Galli, M.~Grassi, D.~Nicolo and F.~Raffaelli {\it et al.},
  arXiv:1301.7225 [physics.ins-det].

\bibitem{Aubert:2009ag} 
  B.~Aubert {\it et al.}  [BaBar Collaboration],
  Phys.\ Rev.\ Lett.\  {\bf 104}, 021802 (2010)
  [arXiv:0908.2381 [hep-ex]].

\bibitem{EDM}
  J. Baron et al. [ACME Collaboration], Science V. 343, N. 6168 (2014) 269; [arXiv:1310.7534].

\bibitem{Bennett:2008dy} 
  G.~W.~Bennett {\it et al.}  [Muon (g-2) Collaboration],
  Phys.\ Rev.\ D {\bf 80}, 052008 (2009)
  [arXiv:0811.1207 [hep-ex]].

\bibitem{Omura} 
  Y.~Omura, E.~Senaha and K.~Tobe,
  arXiv:1502.07824 [hep-ph].

\bibitem{hmutau} 
  D.~Aristizabal Sierra and A.~Vicente,
  Phys.\ Rev.\ D {\bf 90}, 115004 (2014)
  [arXiv:1409.7690 [hep-ph]];
  J.~Heeck, M.~Holthausen, W.~Rodejohann and Y.~Shimizu,
  arXiv:1412.3671 [hep-ph];
  A.~Crivellin, G.~D’Ambrosio and J.~Heeck,
  Phys.\ Rev.\ Lett.\  {\bf 114}, 151801 (2015)
  [arXiv:1501.00993 [hep-ph]];
  L.~de Lima, C.~S.~Machado, R.~D.~Matheus and L.~A.~F.~do Prado,
  arXiv:1501.06923 [hep-ph];
  A.~Crivellin, G.~D’Ambrosio and J.~Heeck,
  Phys.\ Rev.\ D {\bf 91}, no. 7, 075006 (2015)
  [arXiv:1503.03477 [hep-ph]];
  D.~Das and A.~Kundu,
  arXiv:1504.01125 [hep-ph];
  B.~Bhattacherjee, S.~Chakraborty and S.~Mukherjee,
  arXiv:1505.02688 [hep-ph].

\bibitem{Khachatryan:2015kon} 
  V.~Khachatryan {\it et al.}  [CMS Collaboration],
  arXiv:1502.07400 [hep-ex].

\end{thebibliography}
\end{document}